\documentclass{emulateapj}

\usepackage{float}
\usepackage{amsmath}
\shorttitle{Filament Channel Formation}
\shortauthors{Knizhnik et al.}

\newcommand{\beg}[1]{\begin{equation}\label{#1}}
\newcommand{\done}{\end{equation}}
\newcommand{\pd}[2]{\frac{\partial #1}{\partial #2}}

\newcommand{\vecB}{\textbf{B}}
\newcommand{\vecA}{\textbf{A}}

\newcommand{\vecdS}{\textbf{dS}}

\newcommand{\vecn}{\textbf{n}}
\newcommand{\vecr}{\textbf{r}}

\newcommand{\vecv}{\textbf{v}}
\newcommand{\vecx}{\textbf{x}}
\newcommand{\vecy}{\textbf{y}}
\newcommand{\vecz}{\textbf{z}}

\newcommand{\curl}[1]{\nabla\times{#1}}
\newcommand{\divv}[1]{\nabla\cdot{#1}}
\numberwithin{equation}{section}
\begin{document}

\title{Filament Channel Formation via Magnetic Helicity Condensation}

\author{K. J. Knizhnik\altaffilmark{1,2}, S. K. Antiochos\altaffilmark{2}, and C. R. DeVore\altaffilmark{2}}

\altaffiltext{1}{Department of Physics and Astronomy, The Johns Hopkins University, Batimore, Maryland}
\altaffiltext{2}{Heliophysics Science Division, NASA Goddard Space Flight Center, Greenbelt, Maryland}

\begin{abstract}
A major unexplained feature of the solar atmosphere is the accumulation of magnetic shear, in the form of filament channels, at photospheric polarity inversion lines (PILs). In addition to free energy, this shear also represents magnetic helicity, which is conserved under reconnection. In this paper, we address the problem of filament channel formation and show how they acquire their shear and magnetic helicity. The results of 3D simulations using the Adaptively Refined Magnetohydrodynamics Solver (ARMS) are presented that support the model of filament channel formation by magnetic helicity condensation developed by \citet{Antiochos13}. We consider the supergranular twisting of a quasi-potential flux system that is bounded by a PIL and contains a coronal hole (CH). The magnetic helicity injected by the small-scale photospheric motions is shown to inverse-cascade up to the largest allowable scales that define the closed flux system: the PIL and the CH. This process produces field lines that are both sheared and smooth, and are sheared in opposite senses at the PIL and the CH. The accumulated helicity and shear flux are shown to be in excellent quantitative agreement with the helicity-condensation model.  We present a detailed analysis of the simulations, including comparisons of our analytical and numerical results, and discuss their implications for observations. \par


\end{abstract}

\section{Introduction}\label{sec:intro}

Prominences and filaments are among the most stunning features of the solar corona. They are long, thin structures, with lengths of order several hundred Mm, heights of order tens of Mm, and widths of several Mm. Prominences and filaments are known to be the same phenomenon, called by the former name when seen in emission at the solar limb, the latter when seen in absorption on the disk. Filaments form in filament channels \citep{Martin98,Gaizauskas00}, which are located above and concentrated adjacent to polarity inversion lines (PILs), where the radial component of the magnetic field changes sign. Filament channels are strongly sheared magnetic structures that can support substantial mass against solar gravity. A filament comes into existence if sufficient cool plasma accumulates in the coronal magnetic field of the channel. \par

Because filament channels are highly sheared, their nonpotential magnetic fields contain substantial amounts of free energy. This free energy is converted into kinetic and thermal energy of the gas and nonthermal energy of accelerated particles when filaments erupt in coronal mass ejections (CMEs). The shear that is inherent in the filament channels represents not only free energy, but magnetic helicity, which is carried away by the ejected flux ropes formed during CMEs. Filament channels are the only locations in the corona where significant magnetic free energy and helicity are observed. High resolution EUV and X-ray images of the closed-field corona taken by the \emph{Transition Region And Coronal Explorer} (TRACE) mission show a collection of smooth loops everywhere except in these filament channels \citep{Schrijver99}. A fundamentally important question then is: How do filament channels form? \par

Three general classes of mechanisms have been proposed to explain filament channels. One mechanism is flux cancellation \citep{VanB89}. In this model, the coronal magnetic field is sheared by the Sun's large-scale differential rotation. The shear collects and strengthens due to converging flows at the PIL, which cancel and reconnect opposite-polarity fluxes at the photosphere. This reconnection forms low-lying concave-down loops that disappear below the surface and concave-up loops that rise into the corona, generally occurring at multiple locations along the PIL. Thus, flux cancellation invariably produces a twisted coronal flux rope. Many theoretical investigations of filaments assume that the underlying structure is a twisted flux rope \citep[e.g.,][]{Malherbe83, Aulanier98, VanB00}. However, high-resolution observations of filaments \citep{Lin05, Vourlidas10} indicate that filaments are laminar and smooth, showing little evidence of substantial twist. This suggests that flux cancellation, although routinely observed in filament channels \citep[e.g.,][]{Martin98}, is unlikely to be responsible for their development. \par

A second mechanism for filament channel formation is flux emergence, in which a rising sub-photospheric twisted flux rope breaches the surface to produce the filament channel directly. First-principles numerical simulations \citep[e.g.,][]{Fan01, Manchester01, Magara03} show that flux emergence results in the formation of a strongly sheared arcade of coronal loops. The axial field of the flux rope comprises the sheared field of the filament channel, the concave-down portion above forms the overlying quasi-potential arcade, and the concave-up portion below remains submerged beneath the photosphere. The resultant structure has the defining properties of a filament channel, and this mechanism is plausibly responsible for the formation of filament channels in solar active regions. On the other hand, filament channels are observed routinely to form in quiet regions where no significant flux emergence is occurring \citep{Mackay10}. Therefore, it seems probable that flux emergence produces some, but almost certainly not all, solar filament channels.\par

The third mechanism that explains filament channels is the formation of a sheared arcade by localized photospheric flows parallel to the PIL \citep{Antiochos94, DeVore00, 2005ApJ...629.1122D}. The resulting structures are similar to those obtained in the flux-emergence scenario just described, but can be formed in quiet, as well as active, solar regions. Although the requisite shearing flows are observed on the Sun occasionally, they are highly intermittent, and thus seem unlikely to account for the majority of filament channels. \par

A conceptual breakthrough in understanding the formation of filament channels, as well as other large-scale structural features of the solar atmosphere, is the helicity condensation model put forth by 
\citet{Antiochos13}. Any systematic vortical motions associated with the Sun's surface convection, in particular the supergranulation, will inject helicity into the coronal magnetic field. Counter-clockwise rotations in the northern hemisphere, and clockwise in the southern, in agreement with the sense of the solar differential rotation, would inject negative helicity in the north and positive in the south. This pattern of injection is consistent with numerous observations showing that filament channels, filaments, and other structures have a strong preference for left-/right-handed twist in the north/south, also referred to as dextral/sinistral chirality \citep{Martin92, Rust94a, Zirker97, Pevtsov03}. It also is consistent with helioseismological measurements of subsurface flows in supergranules, which reveal that their vorticity is antisymmetric across the equator, with counter-clockwise motions in the north and clockwise in the south \citep{Duvall00, Gizon03, Komm07, Rieutord10}. \par

Although the helicity condensation model asserts that magnetic helicity is injected at supergranulation scales, it does not remain at those scales. Instead, the helicity is transferred to ever larger scales within any unipolar region via magnetic reconnection. Steps 1, 2, $...$ $N$ of this process are illustrated schematically in Figure \ref{fig:condensation}, which shows a coronal hole (`CH') with an adjacent closed-field region, both of positive polarity (`+'), encircled by a polarity inversion line (`PIL'; gray dashed line). Thick black curves are magnetic field lines, which are drawn as dashed curves when on the underside of the magnetic structure to which they belong. The ubiquitous twisting flows of the supergranulation are represented by the black counter-clockwise circular arrows distributed across the surface. At step 1, two neighboring flux tubes (`a' and `b'; light blue) with like-polarity axial fields and similarly twisted azimuthal fields have come into contact. Because the azimuthal fields are anti-parallel at contact points (yellow lines) between tubes, they can reconnect and cancel. The result at step 2 is a newly merged, single flux tube (`a+b'; light purple) containing the combined axial fluxes of the original pair, enclosed by the azimuthal flux that previously wrapped around each individual tube. Helicity is conserved under reconnection in the highly conducting corona, so this merging process has transferred the helicity injected at the supergranulation scale to the next larger scale. This process continues with further mergers, such as that at step 2 between the newly formed, larger flux tube and a neighboring elemental tube (`c'; light blue) at the supergranulation scale. \par

The transfer of helicity to ever-larger scales (`a+b+c+...') continues until the azimuthal flux attains the scales of the encircling PIL and the embedded CH at step $N$. At the PIL, this flux (thick dashed curves with arrowheads, linking the dark purple regions) is trapped on low-lying field lines, where it can only accumulate until it is liberated by a CME. This concentration of azimuthal flux at the PIL, where the helicity is said to `condense,' is precisely the signature of a filament channel. At the CH, in contrast, the azimuthal flux is imparted to high-lying field lines that can open easily into the solar wind, so the field there remains untwisted. Note from the figure that the sense of twist of the azimuthal flux propagating toward the CH (in flux tubes `a', `b', and `c') is clockwise; this is opposite to the sense of twist imparted to the open flux tubes residing within the coronal hole by the counter-clockwise rotations. Thus, the helicity condensation model makes an observational prediction that helicity fluxes measured within the interior of a coronal hole and at its perimeter will have opposite signs \citep{Antiochos13}. \par

An initial numerical investigation of basic predictions of the helicity condensation model has been reported by two of us in \citet{Zhao14}. The ansatz of \citet{Parker72}, in which the corona is modeled by a uniform magnetic field between two horizontal plates, was adopted. A simulation of two neighboring like-polarity, like-twisted flux tubes verified the reconnection and merger of two small tubes into one larger tube, as shown schematically in steps 1 to 2 of our Figure \ref{fig:condensation}. A companion simulation showed that reconnection does not occur if the two flux tubes are twisted in opposite senses: because their azimuthal fields are parallel rather than anti-parallel, the tubes remain separate and distinct instead of merging, as predicted by \citet{Antiochos13}. A second pair of simulations assumed a close-packed, regular hexagonal array of seven rotation cells. In one case, the array was kept spatially fixed through a dozen temporal cycles of turning the flows on and off; in the other, the array of cells was randomly translated and rotated between cycles, to emulate the ever-shifting pattern of the Sun's supergranulation. Both simulations demonstrated that the injected azimuthal flux was transferred via reconnection to the perimeter of the region of flows, again as predicted by the helicity condensation model. Only minor, quantitative differences between the fixed and randomized cases were found, indicating that the evolution of the system is insensitive to the details of the photospheric driving motions so long as the motions are sufficiently complex. For computational simplicity, therefore, we use only a fixed pattern of driving flows in the calculations discussed in this paper. \par

We examine in detail the processes of magnetic helicity injection, transport, and condensation within a plane-parallel corona for a far larger ensemble of rotation cells than that assumed by 
\citet{Zhao14}. In addition, and more fundamentally, we include an interior region that is free of rotation cells, as an elementary model for a coronal hole embedded within a unipolar region of closed field. This revised configuration is a simplified representation of the scenario shown here in Figure \ref{fig:condensation}. Our simulations enable us to test the predictions of the helicity condensation hypothesis with greater generality and complexity than in previous work. We also develop and verify some quantitative, analytic estimates of the early- and late-time behaviors in our simulation. Those results aid our understanding of the plane-parallel system that we have adopted here. They also prepare the way for future studies of more geometrically realistic scenarios having true polarity inversion lines and coronal holes with open fields. \par

The remainder of our paper is organized as follows. In \S \ref{sec:simulations}, we describe the numerical simulation model that we used to test the helicity condensation hypothesis. Our analytical deductions and numerical diagnostics for magnetic twist and helicity are presented in \S \ref{sec:helicity}.  The core of the paper consists of the results and analyses of our numerical simulation results, which are given in \S \ref{sec:Results}. Finally, in \S \ref{sec:discussion} we discuss the implications of our findings for understanding filament channel formation on the Sun. \par

\section{The Numerical Model}\label{sec:simulations}

We use the Adaptively Refined Magnetohydrodynamics Solver \citep[ARMS; e.g.,][]{DeVore08} to solve the equations of magnetohydrodynamics (MHD) in three Cartesian dimensions. 
The equations have the form
\beg{cont}
\pd{\rho}{t}+\divv{\left( \rho\vecv \right)}=0,
\done
\beg{momentum}
\pd{\rho\vecv}{t} + \divv{\left( \rho\vecv\vecv \right)} = - \nabla P + \frac{1}{4\pi} \left( \curl{\vecB} \right) \times \vecB,
\done
\beg{energy}
\pd{T}{t} + \divv{\left( T\vecv \right)} = \left( 2 - \gamma \right) T \divv{\vecv},
\done
\beg{induction}
\pd{\vecB}{t} = \curl{ \left( \vecv \times \vecB \right)}.
\done
Here $\rho$ is mass density, $T$ is temperature, $P$ is thermal pressure, $\gamma$ is the ratio of specific heats, $\vecv$ is velocity, $\vecB$ is magnetic field, and $t$ is time. We close the equations via the ideal gas equation, 
\beg{idealgas}
P=\rho RT,
\done
where $R$ is the gas constant. ARMS employs Flux-Corrected Transport algorithms \citep{DeVore91} and finite-volume representation of the variables to obtain its solutions. Its minimal, but finite, numerical dissipation allows reconnection to occur at electric current sheets associated with discontinuities in the direction of the magnetic field. \par

Our simulation configuration is shown in Figure \ref{fig:initial}. We model the coronal magnetic field as initially straight and uniform ($\vecB = B_0 \hat{\vecx}$) between two plates \citep{Parker72}. Straight flux tubes therefore represent coronal loops, with the apex of each `loop' positioned in the center of the domain. Each of the two boundary plates represents the photosphere. The domain extent 
in $(x,y,z)$ is $[0,L_x]\times[-L_y,+L_y]\times[-L_z,+L_z]$, with $x$ the vertical direction (normal to the photosphere), $L_x=1$, and $L_y=L_z=1.75$. \par

We employ zero-gradient conditions at all times at all six boundaries,
\beg{boundaries}
\begin{split}
\pd{\rho}{n}=0,\\
\pd{T}{n}=0,\\
\pd{\vecv}{n}=0,\\
\pd{\vecB}{n}=0,
\end{split} 
\done
where $n=x,y,z$ is the normal coordinate. The four side boundaries are all open to accommodate lateral expansion of the stressed magnetic field. The top and bottom boundaries are closed, where the magnetic field is line-tied. The footpoints of the field lines move only in response to imposed boundary flows, emulating the slow driving at the dense solar photosphere, rather than in response to coronal magnetic forces. \par

The initial, uniform values used in our dimensionless simulation are $\rho_0=1$, $T_0=1$, $P_0=0.05$, and $B_0=\sqrt{4\pi}$. These choices set the gas constant, $R=0.05$, the Alfv\'en speed, $c_{A0}=B_0 / \sqrt{4\pi\rho_0} = 1$, and the plasma beta, $\beta_0=8\pi P_0/B_0^2=0.1$. The regime $\beta\ll1$ corresponds to a magnetically dominated plasma, which is generally true of the corona. We note that the simulation time therefore is normalized to the time required for an Alfv\'en wave at unit speed ($c_{A0}=1$) to propagate between the top and bottom plates separated by unit distance ($L_x=1$). \par

To model the supergranular twisting of the photospheric footpoints of a finite flux system, we confine the helicity injection to the hexagonal annulus defined by the pattern of surface flows shown in Figure \ref{fig:initial}. The PIL is defined by the outer boundary of this pattern, while the CH is defined by the untwisted field within the inner boundary of the pattern. $N = 84$ identical, circular rotation cells of radius $a_0$ are positioned in the hexagonal array shown, on both the top and bottom plates. Our CH thus has radius $a_c=3a_0$ at the photosphere, while the PIL has radius $a_p=10a_0$. \par

The adaptive mesh refinement capability of ARMS was used to resolve very finely that part of the domain volume where the photospheric flows are imposed and the coronal flux tubes become twisted. Each elemental block of the grid contained $8\times8\times8$ uniform, cubic grid cells. Four such blocks, or 32 grid cells, were used to span the radius $a_0$ of each of our photospheric rotations. The full pattern of rotations, the lanes between them, the coronal hole in the interior, and a buffer region around the outer perimeter of the pattern were covered uniformly by these high-resolution grid blocks. This finely gridded region extended vertically throughout the corona. Outside of this region toward the side walls, the grid was allowed to coarsen by two levels, the grid spacing increasing by a factor of two at each change of refinement level. The resulting grids were about $25\%$ of the sizes of equivalent uniform grids throughout the domain. \par

We set the vertical velocity $v_x\vert_S=0$ at the top and bottom boundaries. The $x$-component of the induction equation (\ref{induction}) can be written
\beg{inductionx}
\pd{B_x}{t}=-(\vecv_\perp\cdot\nabla_\perp)B_x-B_x(\nabla_\perp\cdot\vecv_\perp),
\done
where $\perp$ represents the $y$ and $z$ directions. At $t=0$, $B_x$ is uniform on the boundary, so the first term on the right-hand side vanishes. If the second term also vanishes, because the flows are incompressible, then $\partial B_x /\partial t = 0$ at all times $t$. The latter condition is satisfied identically if, for each rotation cell,
\beg{flowv}
\vecv_\perp = \hat{\vecx} \times \nabla \chi(r,t)
\done
where $\chi$ is any scalar function, here taken to depend only upon time $t$ and the radial coordinate $r$ centered on the rotation cell. We set 
\beg{flowchi}
\chi(r,t)=\Omega_0 a_0^2 f(t) g(r).
\done
The temporal profile $f(t)$ is given by
\beg{foft}
f(t) = \frac{1}{2} \left[ 1 - \cos \left( 2\pi \frac{t}{\tau} \right) \right],
\done
so that the flows ramp up from zero and then back down to zero over the period $\tau$; the corresponding displacement is proportional to the integral of $f(t)$,
\beg{foftI}
F(t) = \frac{1}{2} \left[ t - \frac{\tau}{2\pi} \sin \left( 2\pi \frac{t}{\tau} \right) \right].
\done
The spatial profile $g(r)$ is given by
\beg{gofr}
g(r)=\frac{1}{6}\left[1-\left(\frac{r}{a_0}\right)^{6}\right] - \frac{1}{10}\left[1-\left(\frac{r}{a_0}\right)^{10}\right].
\done
Outside of $r=a_0$, we fix $g(r)=0$. With this form of $\chi(r,t)$, the angular rotation rate is given by
\beg{omega}
\begin{split}
\Omega(r,t)&=\Omega_0 a_0^2 f(t) \frac{1}{r}\frac{dg}{dr}\\
&=- \Omega_0 f(t) \left(\frac{r}{a_0}\right)^4 \left[1-\left(\frac{r}{a_0}\right)^4\right]
\end{split}
\done
for $r \le a_0$. For the flow parameters, we choose $a_0=0.125$, $\Omega_0=7.5$, and $\tau=3.35$. These set $\vert \vecv_\perp \vert_{\rm max}=0.200$ and a maximum angle $\vert \Delta \phi \vert_{\rm max} = \pi$ of the clockwise rotation within each cell over the period $\tau$ of each cycle. \par

\section{Magnetic Helicity and Twist}\label{sec:helicity}
\subsection{Magnetic Helicity Injection}\label{sec:helinj}
In a volume $V$ bounded by magnetic flux surfaces $S$, so that $\vecB\cdot\hat{\vecn}|_S=0$ with $\hat{\vecn}$ the unit vector normal to $S$, the magnetic helicity is defined simply as 
\beg{helicity}
H=\int_V{\vecA\cdot\vecB \; dV,}
\done
where $\vecB=\curl{\vecA}$ and $\vecA$ is the vector potential \citep{Berger99}. This integral measures the general topological property of field line linkages. In the corona and in our simulation domain, where magnetic field lines enter from or exit to the photosphere, the bounding surface is not a flux surface, and the relative magnetic helicity \citep{1984JFM...147..133B} must be used in place of Equation 
(\ref{helicity}).  We adopt the gauge-invariant form of \citet{Finn85}, 
\beg{relativehelicity}
H=\int_V{(\vecA+\vecA_P)\cdot(\vecB-\vecB_P) \; dV},
\done
where $\vecB_P = \curl{\vecA_P}$ is a current-free field ($\curl{\vecB_P} = 0$) satisfying $\vecB_P\cdot\hat{\vecn}|_S=\vecB\cdot\hat{\vecn}|_S$. It can be shown through integrations by parts that, under ideal evolution, the time derivative of Equation (\ref{relativehelicity}) leads to 
\beg{dHdt}
\frac{dH}{dt}=2\oint_S{\left[ \left( \vecA_P \cdot \vecv \right) \vecB - \left( \vecA_P \cdot \vecB \right) \vecv \right] \cdot \vecdS.}
\done
The first term represents the effects of twisting or shearing motions on the boundary, while the second term represents helicity injected (removed) by the emergence (submergence) of helical field across the boundary. Thus, the only contributions to $dH/dt$ are due to motions on or through the bounding surface $S$. Absent such motions, the helicity $H$ is conserved perfectly in the volume $V$. In highly conducting plasmas that are almost ideal, the helicity is conserved even in the presence of a small localized resistivity that gives rise to magnetic reconnection \citep{Woltjer58,Taylor74,1986RvMP...58..741T,Berger84b}. \par

To monitor the magnetic helicity in our simulations, we calculate the volume integral and surface injection rate using Equations (\ref{relativehelicity}) and (\ref{dHdt}), respectively. The vector potential for the uniform, current-free initial field is
\beg{potA}
\vecA_p = \frac{B_0}{2} (y\hat{\vecz}-z\hat{\vecy}),
\done 
so that $\vecB_p = \nabla \times \vecA_p = B_0 \hat{\vecx}$. Using $\vecA_p$ as the boundary value at the photosphere at $x=0$, and adopting the gauge $A_x=0$, the vector potential at all times $t$ and heights $x$ becomes 
\beg{vecpotA}
\begin{split}
\vecA(x,y,z,t) = \vecA_p(y,z) &+ \hat{\vecy}\int_0^x {dx'B_z(x',y,z,t)}\\
&- \hat{\vecz}\int_0^x {dx'B_y(x',y,z,t).}
\end{split}
\done
This expression was used to evaluate the volume integral (\ref{relativehelicity}) for $H$. Because the helicity injected by each rotational cell is independent of its position on either plate, we can evaluate Equation (\ref{dHdt}) for the helicity injection rate due to a single flux tube centered at $(x,y,z)=(0,0,0)$. We rewrite Equation (\ref{potA}) as 
\beg{vecAcyl}
\vecA_p=\frac{B_0}{2}r\hat{\phi},
\done
with $r$ the radial coordinate centered on the cell. Because there is no motion through the boundary, Equation (\ref{dHdt}) simplifies to 
\beg{calcdhdt}
\begin{split}
\frac{dH'}{dt} &= - \int_{S'} {r \hat{\phi} \cdot \vecv B_0^2 dS} \\
&= - \int_{S'} {r^2 \Omega(r,t) B_0^2 dS},
\end{split}
\done
where $H'$ is the helicity contributed by the footprint $S'$ of the flux tube, and we have used $\vecv\cdot\hat{\phi} = r\Omega(r,t)$. Substituting Equation (\ref{omega}) for $\Omega$, performing the area integral, and doubling the result to include the rotation cells at both the top and bottom plates yields for the helicity injection rate per flux tube 
\beg{dhdtanalytic}
\frac{dH_{f}'}{dt} = + \frac{\pi}{6} \Omega_0 a_0^4 B_0^2 f(t).
\done
Integrating (\ref{dhdtanalytic}) using (\ref{foftI}), we obtain for the time history of the helicity contributed per flux tube 
\beg{hanalytic}
\begin{split}
H_{f}'(t) &= \frac{\pi}{12} \Omega_0 a_0^4 B_0^2 \left[t - \frac{\tau}{2\pi} \sin \left(2\pi\frac{t}{\tau}\right) \right] \\
&= \langle \frac{dH_{f}'}{dt} \rangle \left[t - \frac{\tau}{2\pi} \sin \left(2\pi\frac{t}{\tau}\right) \right].
\end{split}
\done
The average rate of helicity injection and the resultant injected helicity per flux tube over one cycle of duration $\tau$ are 
\beg{avgdHdt}
\langle \frac{dH_{f}'}{dt} \rangle = \frac{\pi}{12} \Omega_0 a_0^4 B_0^2 = 6.0 \times 10^{-3}
\done
and 
\beg{changeH}
\Delta {H_{f}'} = \frac{\pi}{12} \Omega_0 a_0^4 B_0^2 \tau = 2.0 \times 10^{-2},
\done
respectively, after substituting the numerical values of the parameters given previously. \par 

\begin{figure*}[!p]
\centering\includegraphics[scale=0.55]{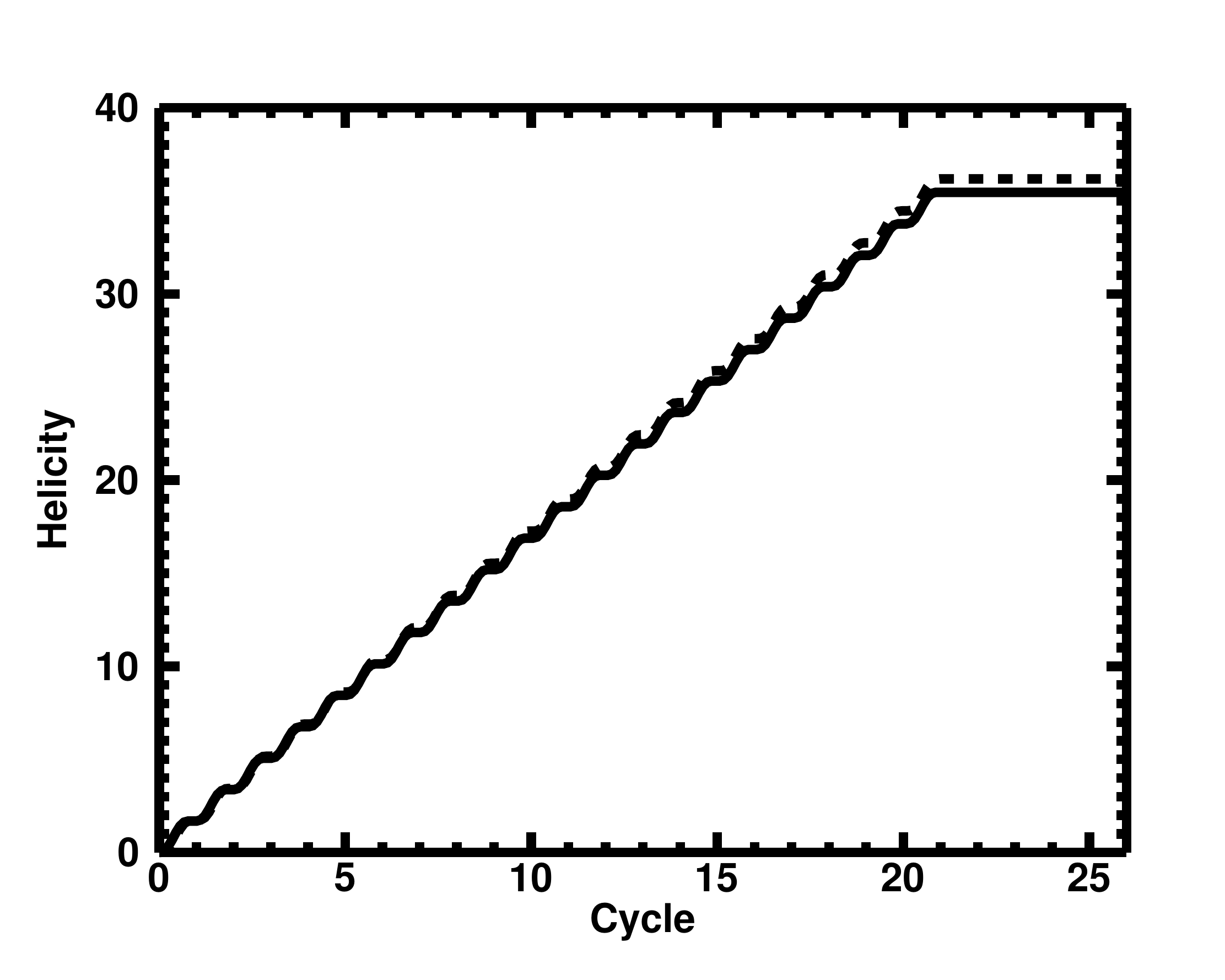}
\caption{Accumulated magnetic helicity in the domain as calculated numerically (solid line) from Equation (\ref{relativehelicity}) and analytically (dashed line) from Equation (\ref{heltime}), demonstrating magnetic helicity conservation to a high degree of accuracy. 
\label{fig:heldiag}}
\end{figure*}

We performed 21 cycles of duration $\tau$ of the flows, each of which twists the field by a maximum angle of $\pi$ on both the top and bottom boundaries, thus yielding up to one full rotation within each flux tube. At the end of the 21 twisting cycles, we performed 5 more cycles during which the field was allowed to relax, i.e., we fixed ${\vecv_\perp}\vert_S=0$, so no new helicity was injected into the system. During the 21 twisting cycles, we can multiply Equations (\ref{avgdHdt}) and (\ref{changeH}) by the $N=84$ flux tubes in the hexagonal pattern to obtain the total average helicity injection rate and helicity injected per cycle, 
\beg{gvals}
\begin{split}
\langle \frac{dH}{dt} \rangle &= 5.0 \times 10^{-1}, \\
\Delta {H} &= 1.7 \times 10^{0}.
\end{split}
\done
Figure \ref{fig:heldiag} compares the numerical results from the volume integral in Equation (\ref{relativehelicity}) with the theoretical prediction obtained by multiplying Equation (\ref{hanalytic}) by 
$N=84$,
\beg{heltime}
H(t) = 5.0 \times 10^{-1} \left[t - \frac{\tau}{2\pi} \sin \left(2\pi\frac{t}{\tau}\right) \right],
\done
for 21 cycles. Clearly, the numerically calculated helicity matches the analytically calculated value to high accuracy. This is a very important diagnostic, because conservation of magnetic helicity under reconnection is crucial to the helicity condensation model. The agreement between the numerically and analytically calculated helicities demonstrates that we are calculating accurately the transport of helicity and twist flux throughout the domain. The algorithms employed by ARMS conserve magnetic flux to machine accuracy, so that the minimal numerical diffusion allows the twist flux to spread out without decreasing. As a result, evidently, a slightly larger axial flux is enclosed by the twist flux, so that the numerically calculated helicity is slightly larger than the analytic value. In any case, the figure demonstrates conclusively that, in spite of the large amount of reconnection that occurs inside the volume, the only significant change in magnetic helicity is due to the twisting motions imposed on the boundaries. \par

\subsection{Twist Flux Generation}\label{sec:twist}

For our physical system we expect that all the helicity is due to the linkages between the uniform background field and the twist flux generated by the photospheric motions. Consequently, in order to understand the helicity evolution, it is highly instructive to follow the evolution of the twist flux.  In fact, the surface helicity integral, Equation (\ref{calcdhdt}), can be recast as a relationship between the rates of helicity injection and twist-flux generation. This result is useful within the individual flux tubes formed early in our simulation and, in addition, when applied across the entire hexagonal array of rotation cells during the late stages of evolution to measure the helicity condensation. 

Over the time interval $dt$, rotation at the rate $\Omega(r,t)$ conveys the vertical field $B_0$ along a strip of length $r \Omega(r,t) dt$ at radius $r$ through the base of the plane $\phi={\rm constant}$. This motion increments the twist flux per unit length in $r$ that passes through that plane, $\Psi_{tw}(r,t)$, by the amount of axial flux that is conveyed across its base,
\beg{dPsitw}
\frac{\partial \Psi_{tw}(r,t)}{\partial t} = - r \Omega(r,t) B_0. 
\done
The total change of the (net) twist flux across the footprint $S'$ of the rotational motion is the radial integral 
\beg{dPhitw}
\frac{d\Phi_{tw}}{dt} = \int_{S'} {dr \; \frac{\partial \Psi_{tw}(r,t)}{\partial t}}.
\done
We note that the minus sign in (\ref{dPsitw}) ensures that a clockwise rotation ($\Omega < 0$) of a positive vertical field ($B_0 > 0$) induces a positive change in the twist flux ($d\Psi_{tw} / dt > 0$). This increment in the twist flux wraps around the axial (vertical) flux that is enclosed within radius $r$,
\beg{Psiax}
\Psi_{ax}(r) = 2\pi \int_0^r {dr' \; r' B_0} = \pi r^2 B_0.
\done
Using (\ref{dPsitw}) and (\ref{Psiax}), the surface helicity integral (\ref{calcdhdt}) therefore can be rewritten 
\beg{calcdhdt3}
\frac{dH'}{dt} = 2 \int_{S'} {dr \; \Psi_{ax}(r) \frac{\partial \Psi_{tw}(r,t)}{\partial t}}.
\done
In this form, magnetic helicity as a measure of the linkages between the axial and twist fluxes is explicit. By rewriting the twist flux increment (\ref{dPsitw}) in terms of the corresponding axial flux increment using (\ref{Psiax}),
\beg{dPsitw2}
\frac{\partial \Psi_{tw}(r,t)}{\partial t} = - \frac{\Omega(r,t)}{2\pi} \frac{d\Psi_{ax}(r)}{dr},
\done
the helicity integral (\ref{calcdhdt3}) can be recast further into the alternate expression 
\beg{calcdhdt4}
\frac{dH'}{dt} = - \frac{1}{\pi} \int_{S'} {dr \; \Omega(r,t) \Psi_{ax}(r) \frac{d\Psi_{ax}(r)}{dr}}. 
\done
This version can be integrated immediately for any axisymmetric flux distribution, for the special case of rigid-body rotation ($\Omega$ independent of $r$). \par

The rate of twist-flux generation in a single flux tube is calculated by evaluating the integral (\ref{dPhitw}). Substituting from Equation (\ref{omega}) for $\Omega$ into (\ref{dPsitw}), and doubling the result to include the contributions from both plates, we obtain 
\beg{dtwdtanalytic}
\frac{d\Phi_{tw,f}}{dt} = +\frac{2}{15} \Omega_0 a_0^2 B_0 f(t).
\done
The corresponding average rate of twist flux injection and the resultant twist flux injected per flux tube over one cycle of duration $\tau$ are 
\beg{avgdtwdt}
\langle \frac{d\Phi_{tw,f}}{dt} \rangle = \frac{1}{15} \Omega_0 a_0^2 B_0 = 2.8 \times 10^{-2}
\done
and
\beg{changetw}
\Delta \Phi_{tw,f} = \frac{1}{15} \Omega_0 a_0^2 B_0 \tau = 9.3 \times 10^{-2},
\done
respectively, after substituting the numerical values of the parameters. \par

The surface helicity integral in the form (\ref{calcdhdt3}) can be used to relate the rate of helicity injection to the rate of generation of large-scale twist flux by reconnection. We make the approximation that the twist flux is concentrated into one or more bands of radial extent $\left[ r_{i_-}, r_{i_+} \right]$ within which $\Omega$ is quasi-uniform. Note that for the large-scale twist flux, 
$\Omega$ is only an effective rotation due to reconnection, not a true rotation due to mass flow.  Equations (\ref{dPhitw}) and (\ref{dPsitw2}) then yield 
\beg{dPhitw2}
\begin{split}
\frac{d\Phi_{tw,r_i}}{dt} &= - \frac{1}{2\pi} \int_{r_{i_-}}^{r_{i_+}} {dr \; \Omega(r,t) \frac{d \Psi_{ax}(r)}{dr} dr} \\
&= - \frac{1}{2\pi} \langle \Omega \rangle_{i} \left[ \Psi_{ax} \left( r_{i_+} \right) - \Psi_{ax} \left( r_{i_-} \right) \right]
\end{split}
\done
for the rate of twist-flux generation in the $i$th band, with $\langle \Omega \rangle_{i}$ the average angular rotation rate in the band. The associated helicity generation rate, from Equation (\ref{calcdhdt4}), is 
\beg{calcdhdt5}
\begin{split}
\frac{dH'_{r_i}}{dt} &= - \frac{1}{\pi} \int_{r_{i_-}}^{r_{i_+}} {dr \; \Omega(r,t) \Psi_{ax}(r) \frac{d\Psi_{ax}(r)}{dr}} \\
&= - \frac{1}{2\pi} \langle \Omega \rangle_{i} \left[ \Psi_{ax}^2 \left( r_{i_+} \right) - \Psi_{ax}^2 \left( r_{i_-} \right) \right] \\
&= + 2 \langle \Psi_{ax} \rangle_{i} \frac{d\Phi_{tw,r_i}}{dt},
\end{split}
\done
where the average axial flux encircled by the band is 
\beg{avgpsiax}
\langle \Psi_{ax} \rangle_{i} = \frac{1}{2} \left[ \Psi_{ax} \left( r_{i_+} \right) + \Psi_{ax} \left( r_{i_-} \right) \right].
\done
We can substitute the numerical rates of helicity and twist-flux generation for a single flux tube, from (\ref{avgdHdt}) and (\ref{avgdtwdt}) respectively, into (\ref{calcdhdt5}) to determine $\langle 
\Psi_{ax} \rangle_{i}$. The procedure yields an effective radius $r_i \approx 0.8a_0 = 0.10$ where the axial flux $\langle \Psi_{ax} \rangle_{i}$ is evaluated; this location is very near the peak angular rotation within the individual flux tubes. For application to the late-time flux distribution in our simulations, we now sum the contributions from (\ref{calcdhdt5}) for multiple bands, obtaining 
\beg{dhdttotal}
\frac{dH}{dt} = \sum_i \frac{dH'_{r_i}}{dt} = 2 \sum_i \langle \Psi_{ax} \rangle_{i} \frac{d\Phi_{tw,r_i}}{dt}.
\done
This relationship links the total rate of helicity injection into the domain, on the left-hand side, to the twist-flux generation rates in the bands, mediated by the enclosed axial fluxes, on the right-hand side. \par

\section{Results}\label{sec:Results}

\subsection{Twist Flux Condensation}\label{sec:condensation}

\begin{figure*}[!p]
\centering\includegraphics[scale=0.75,trim=0.3cm 8.5cm 0.3cm 8.5cm, clip=true]{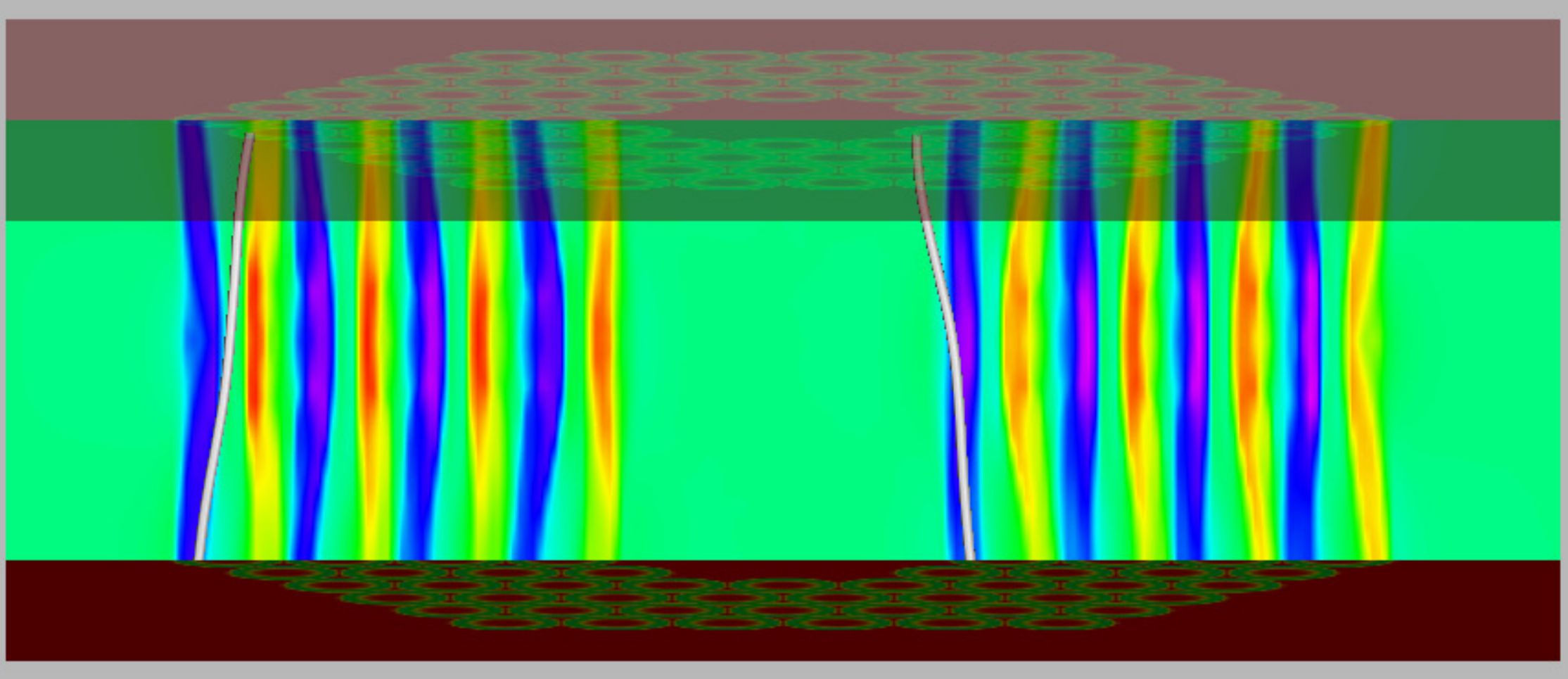}
\centering\includegraphics[scale=0.67]{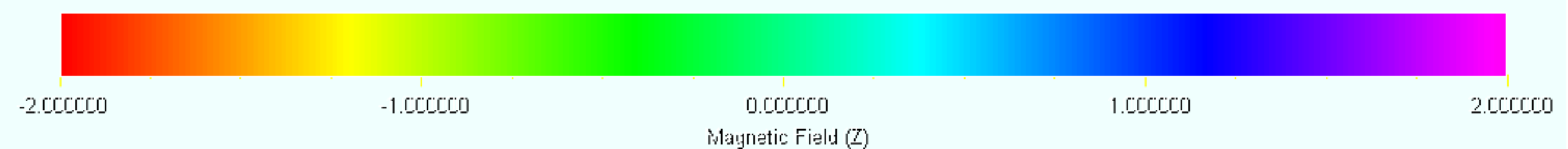}
\centering\includegraphics[scale=0.75,trim=0.3cm 8.5cm 0.3cm 8.5cm, clip=true]{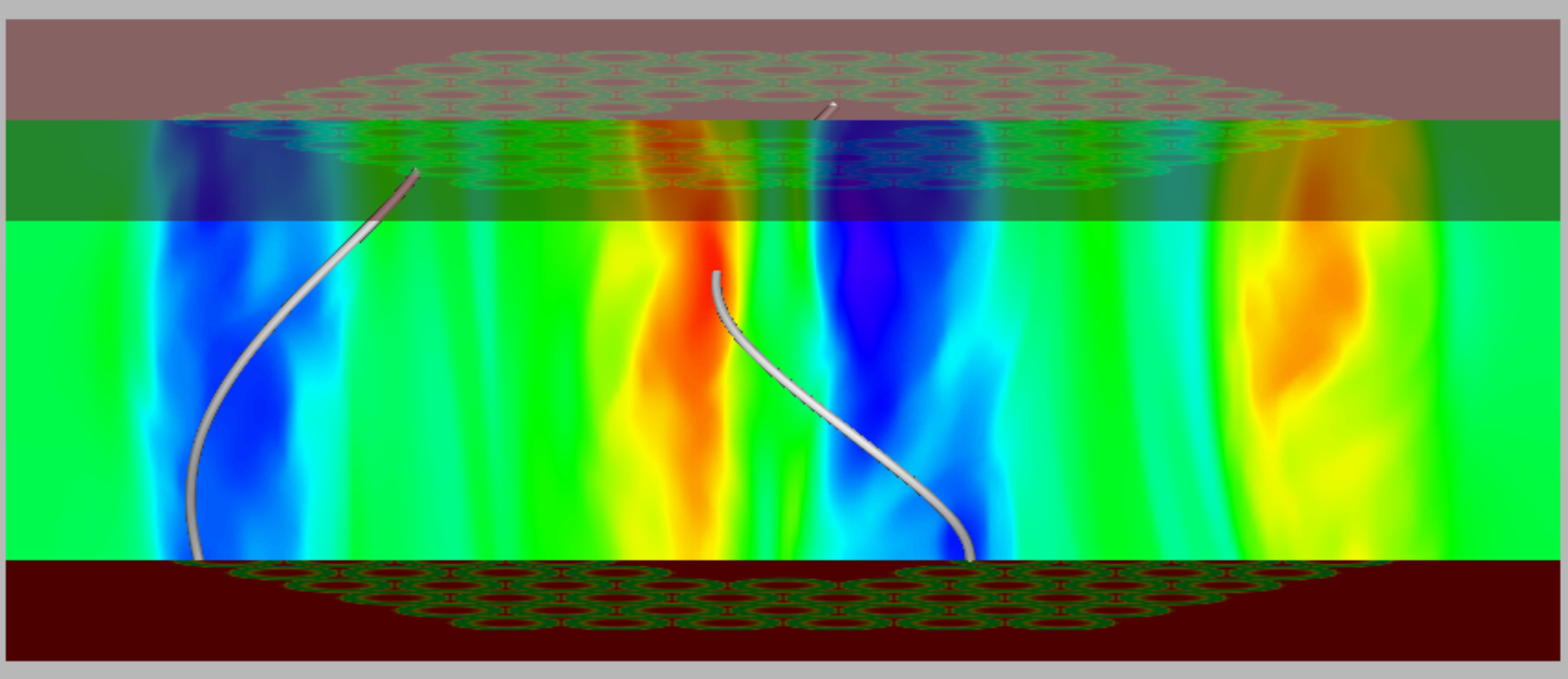}
\centering\includegraphics[scale=0.67]{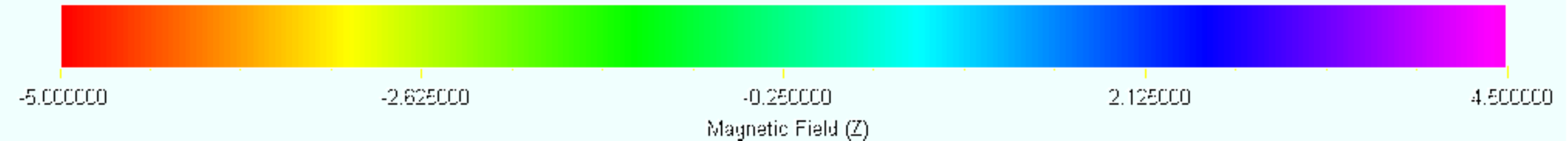}
\caption{Horizontal field $B_z$ (color shading) through the $z = 0$ vertical plane between the top and bottom boundaries. Red/yellow (blue/teal) represents field pointing into (out of) the plane. Two field lines (white) indicate how much twist is present. Top: After 1 twist cycle. Bottom: After 21 twist cycles and 5 relaxation cycles.
\label{fig:fluxandfieldlines}}
\end{figure*}

We begin the presentation of our results by describing qualitatively the global evolution of the system. As will be seen, the magnetic field transitions from numerous individual, twisted flux tubes at early times to a configuration dominated at late times by large-scale flux accumulations at the perimeter of the hexagonal annulus where the rotational flows are imposed. The outer boundary of the flux system, the PIL, is demarcated by the transition to the flow-free region outside the annulus; the inner boundary, the CH, consists of the flow-free region in the center of the annulus. In Figure \ref{fig:fluxandfieldlines}, we show the $y$ component of the magnetic field, $B_y$, on a vertical plane cut through the center of the domain, along with two field lines positioned at the PIL and CH boundaries, at two times in the simulation. After one twist cycle (top), every pair of rotation cells on the top and bottom planes is the footprint of an associated flux tube whose $y$ component of magnetic field goes into and comes out of the vertical plane. The field lines contain slightly less than one full turn of twist. The field line at the PIL, on the left side of the image, is confined within one flux tube; the field line at the CH, on the right side, in contrast, has reconnected already, linking one rotation cell at the bottom boundary to the cell next to its partner at the top.  After 21 twist cycles and 5 relaxation cycles (bottom), the twist flux has condensed almost completely to the PIL and CH boundaries. The field line at the right wraps about three-fourths of the way around the CH, crossing about 10 rotation cells; the field line at the left wraps just under one-sixth of the way around the PIL, crossing only four cells. \par

In Figure \ref{fig:early} we plot the twist component of the magnetic field, i.e., the azimuthal ($\hat{\phi}$) component, $B_\phi$, at azimuthal angles $\phi=0^\circ$, $30^\circ$, $45^\circ$, $60^\circ$, and $90^\circ$ after one Alfv\'en time, $t=1$. Red/yellow (blue/teal) contours show where the field points in the $+\hat{\phi}$ ($-\hat{\phi}$) direction. At this early time, the twist field is concentrated along the boundaries of the individual flux tubes. At the contact point between any two adjacent tubes, all of which are twisted in the same sense, the twist field switches sign. Due to the nonaxisymmetric layout of the velocity pattern, the number of contact points varies with the angle at which the cut is taken. The cut at $\phi=0^\circ$ passes through the centers of two flux tubes between the CH and the PIL on both sides, the cuts at $\phi=30^\circ$ and $90^\circ$ pass through the centers of four tubes, and the cuts at $\phi=45^\circ$ and $60^\circ$ pass through the centers of two tubes and parts of two others. \par

Figure \ref{fig:late} shows the same angle cuts after 21 twist cycles and 5 relaxation cycles. By this time, magnetic reconnection between flux tubes has allowed the magnetic helicity to condense at the boundaries of the system -- the PIL and the CH, as shown schematically in Figure \ref{fig:condensation}. Note that, regardless of how many rotation cells are bisected by these cuts, the final distribution of twist field contains just two bands of opposite polarity between, and concentrated near, the PIL and the CH. All of these features are consistent with the helicity condensation model \citep{Antiochos13}. \par

\subsection{Spatial Distribution}\label{sec:distribution}

To understand better the quantitative features of our results, we now develop some measures of the spatial distribution of the twist flux through the half plane $y=0$, $z\ge0$. In Figure \ref{fig:signedflx} we plot the signed twist fluxes, $\Phi_{tw}^+$ (solid curve) and $\Phi_{tw}^-$ (dashed curve), 
\beg{twistfluxes}
\begin{split}
{\Phi_{tw}^+} &= \int_0^{L_x} dx \int_0^{L_z} dz \; {B_{tw}^+}, \\
{\Phi_{tw}^-} &= \int_0^{L_x} dx \int_0^{L_z} dz \; {B_{tw}^-},
\end{split}
\done
and their sum, the net signed twist flux, $\Phi_{tw} = \Phi_{tw}^+ + \Phi_{tw}^-$ (dotted curve). The corresponding signed twist fields are defined by 
\beg{twistfields}
\begin{split}
B_{tw}^+ &= \frac{1}{2} \left( B_\phi + \vert B_\phi \vert \right) \ge 0, \\
B_{tw}^- &= \frac{1}{2} \left( B_\phi - \vert B_\phi \vert \right) \le 0.
\end{split}
\done
In this particular half plane, $B_\phi = B_y$. At all times, there are almost exactly equal amounts of positive and negative twist flux, owing to the symmetry of the imposed rotational flows and the uniformity of the vertical magnetic field, as we now show. By integrating the $y$ component of the induction equation, we obtain for the net signed twist flux, $\Phi_{tw}$, the result 
\beg{nettwistflux}
\begin{split}
\frac{d\Phi_{tw}}{dt} &= \int_0^{L_x} dx \int_0^{L_z} dz \; {\frac{\partial B_y}{\partial t}} \\
&= \int_0^{L_z} dz \; \left[ v_y B_x - v_x B_y  \right]_{x=0}^{x=L_x} \\
&\phantom{=} \; + \int_0^{L_x} dx \; \left[ v_y B_z - v_z B_y \right]_{z=0}^{z=L_z} \\
&\approx \int_0^{L_z} dz \; \left[  v_y B_x \right]_{x=0}^{x=L_x} \\
&= -2 \int_0^{L_z} dz \; \left[ v_y B_x \right]_{x=0} \\
&= 0.
\end{split}
\done
The $v_x$ terms vanish identically at the top and bottom $x$ boundaries, while the $v_y$ and $v_z$ terms vanish approximately at the inner and outer $z$ boundaries. Only the $v_y B_x$ integrals remain. 
In those integrals, $B_x$ is uniform, and although $v_y$ is antisymmetric at the top and bottom $x$ boundaries, it averages to zero along $z$. Hence, the net twist flux approximately vanishes. \par

\begin{figure*}[!p]
\centering\includegraphics[scale=0.65, trim=1.0cm 6cm 0.75cm 6cm, clip=true]{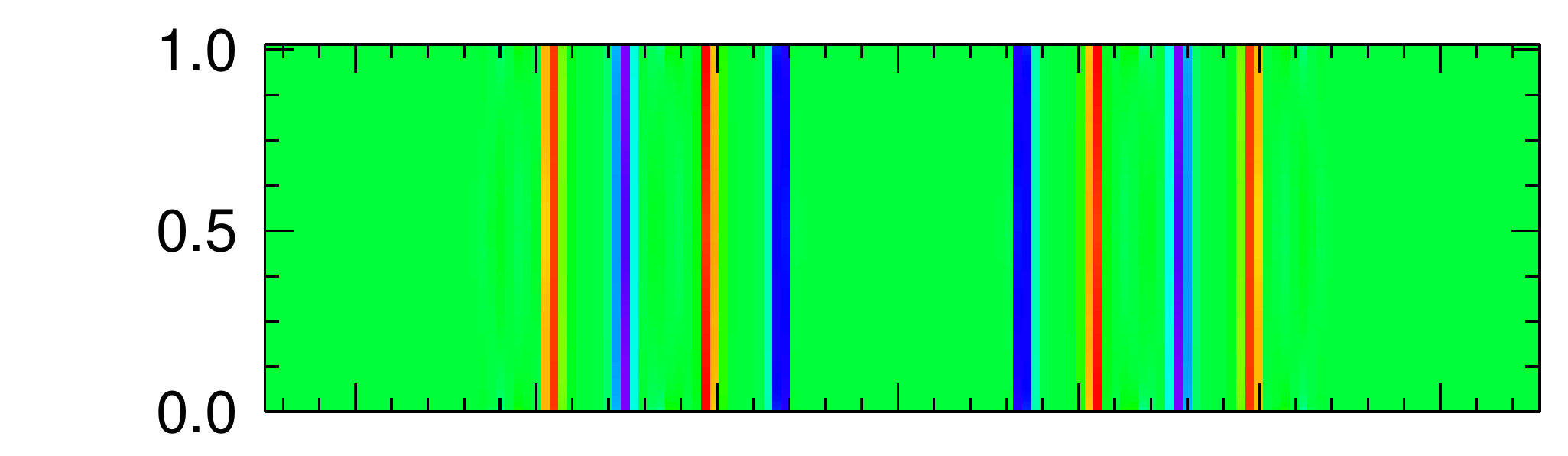}
\centering\includegraphics[scale=0.65, trim=1.0cm 6cm 0.75cm 6cm, clip=true]{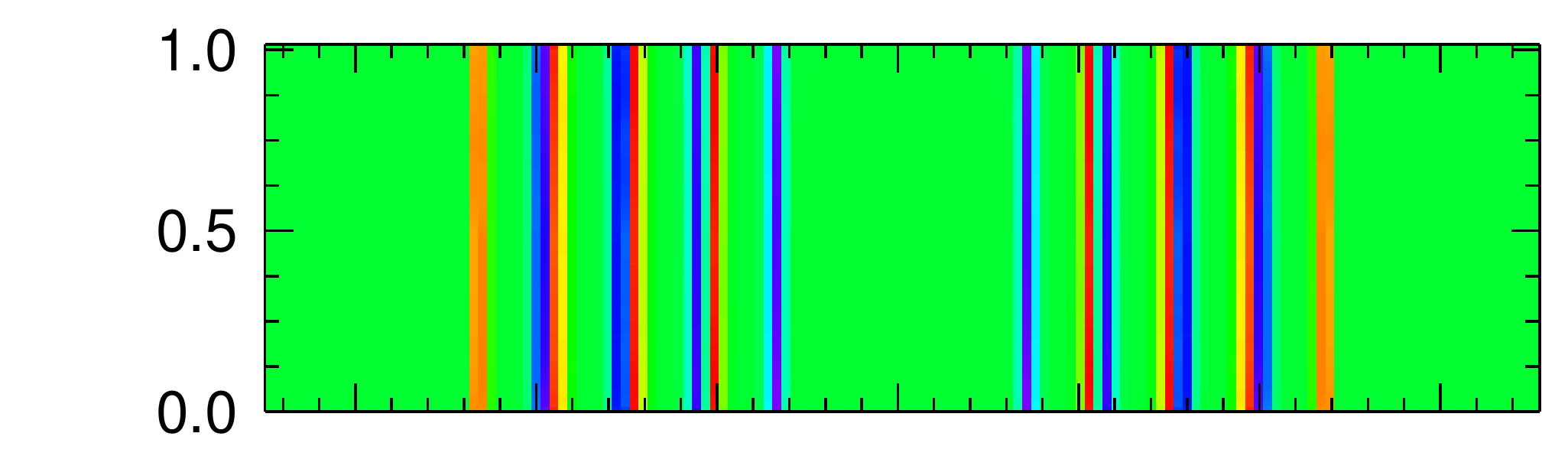}
\centering\includegraphics[scale=0.65, trim=1.0cm 6cm 0.75cm 6cm, clip=true]{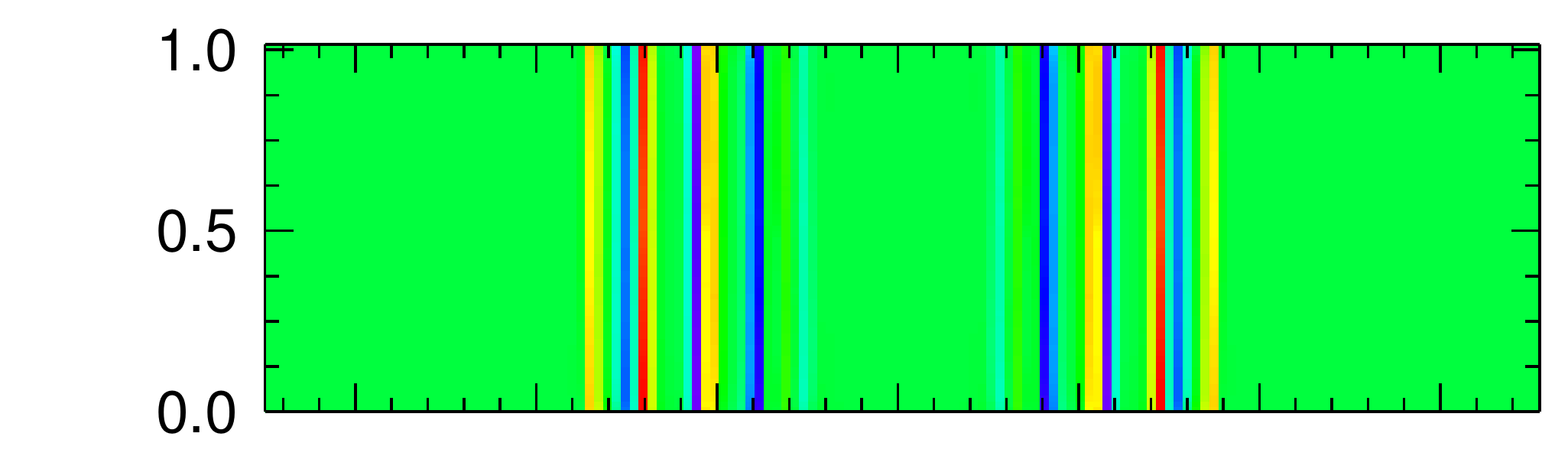}
\centering\includegraphics[scale=0.65, trim=1.0cm 6cm 0.75cm 6cm, clip=true]{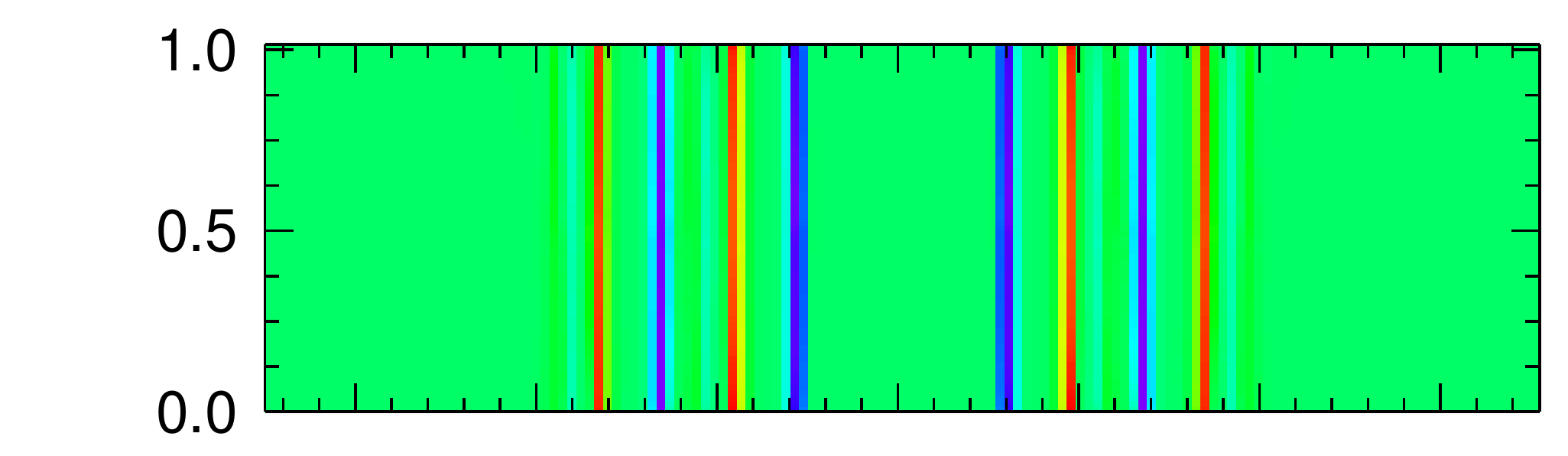}
\centering\includegraphics[scale=0.65, trim=1.0cm 3cm 0.75cm 6cm, clip=true]{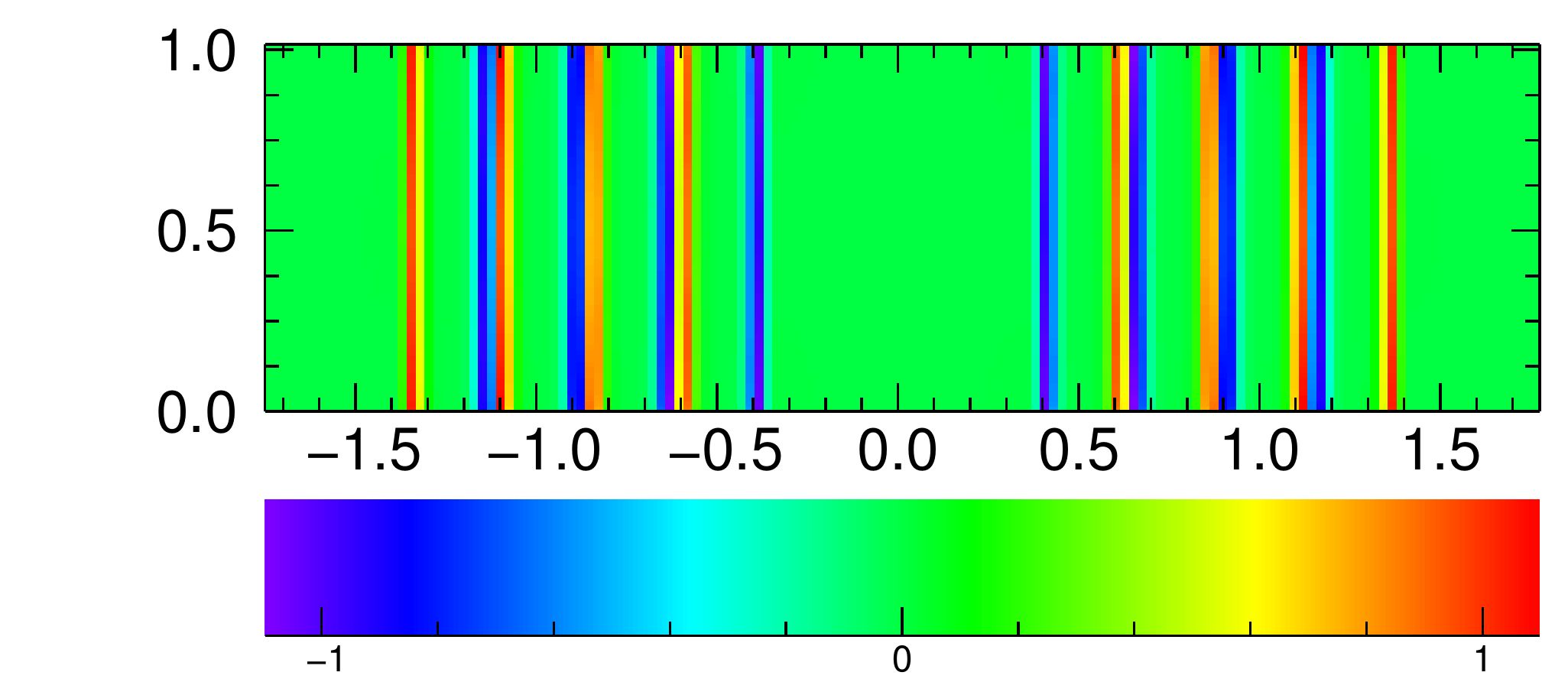}
\caption{Vertical plane cuts of the twist component of the magnetic field, $B_\phi$ (color shading), after one Alfv\'en time ($t = 1$) at angles $\phi=$ $0^\circ$, $30^\circ$, $45^\circ, 60^\circ$, and $90^\circ$ (with respect to the $y$ axis). Red/yellow (blue/teal) represents field pointing in the $+\hat{\phi}$ ($-\hat{\phi}$) direction. At this early time, each pair of rotation cells hosts a clearly delineated individual flux tube.
\label{fig:early}}
\end{figure*}

\begin{figure*}[!p]

\centering\includegraphics[scale=0.65, trim=1.0cm 6cm 0.75cm 6cm, clip=true]{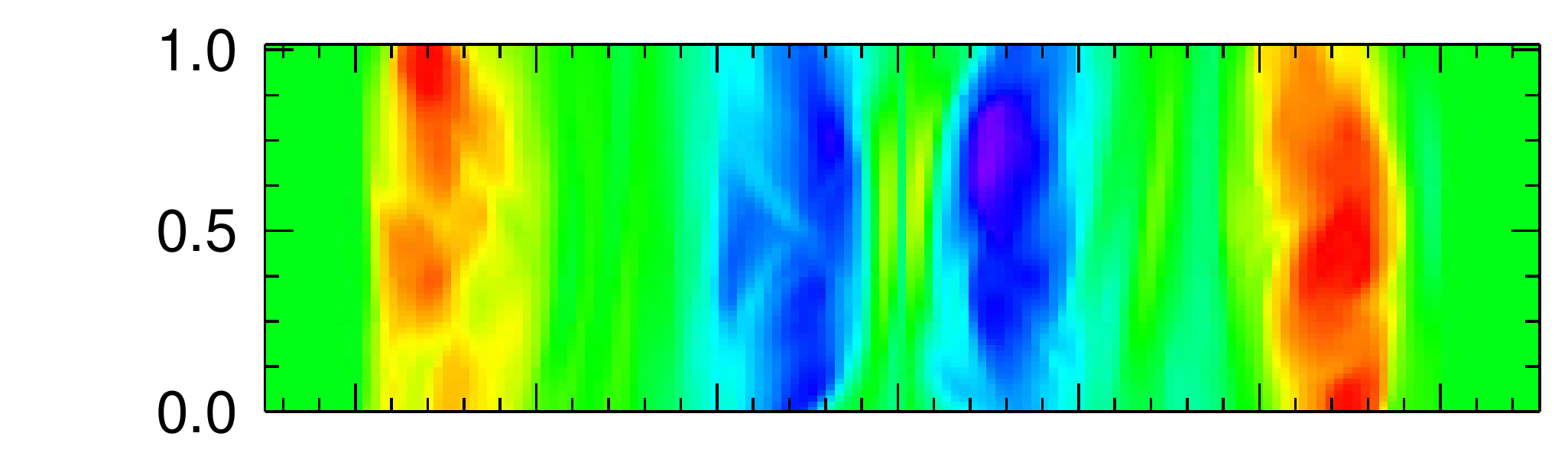}
\centering\includegraphics[scale=0.65, trim=1.0cm 6cm 0.75cm 6cm, clip=true]{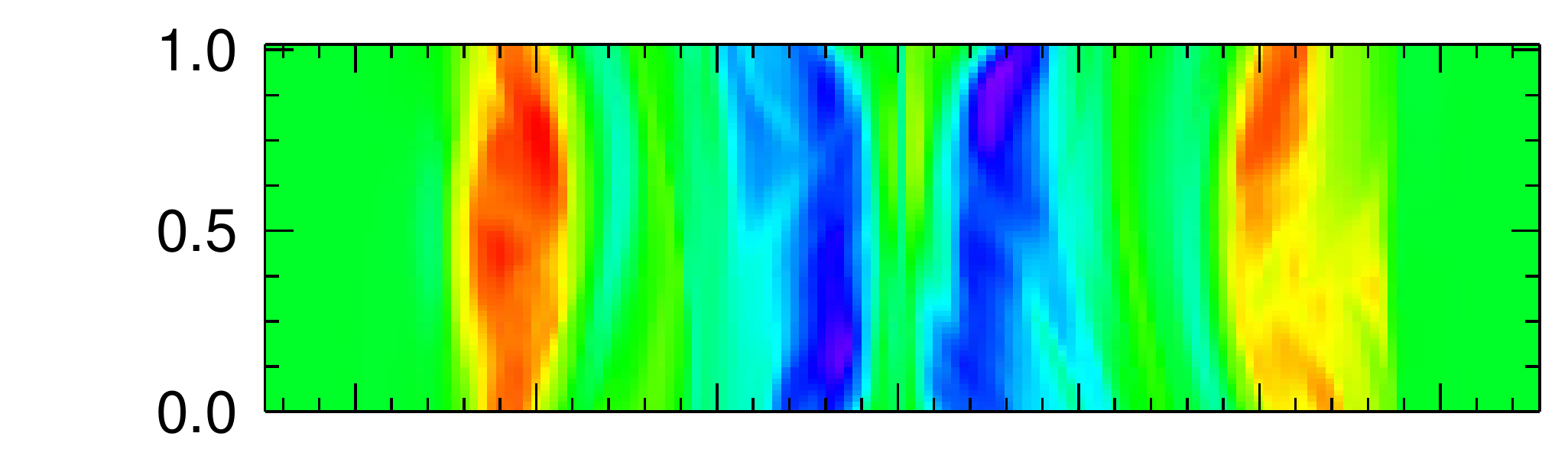}
\centering\includegraphics[scale=0.65, trim=1.0cm 6cm 0.75cm 6cm, clip=true]{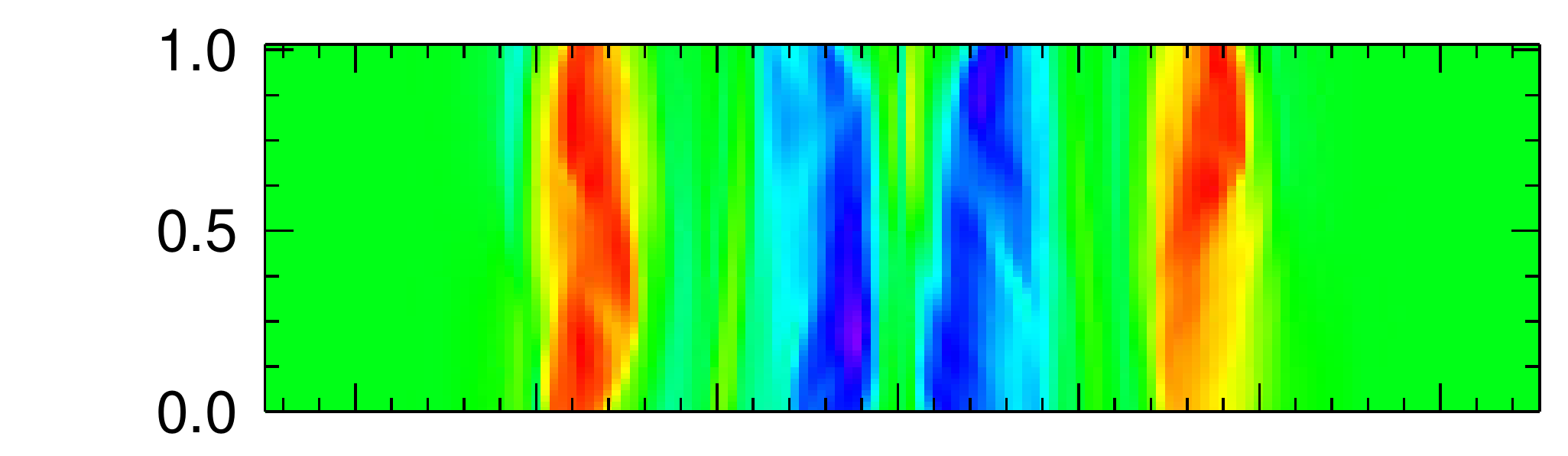}
\centering\includegraphics[scale=0.65, trim=1.0cm 6cm 0.75cm 6cm, clip=true]{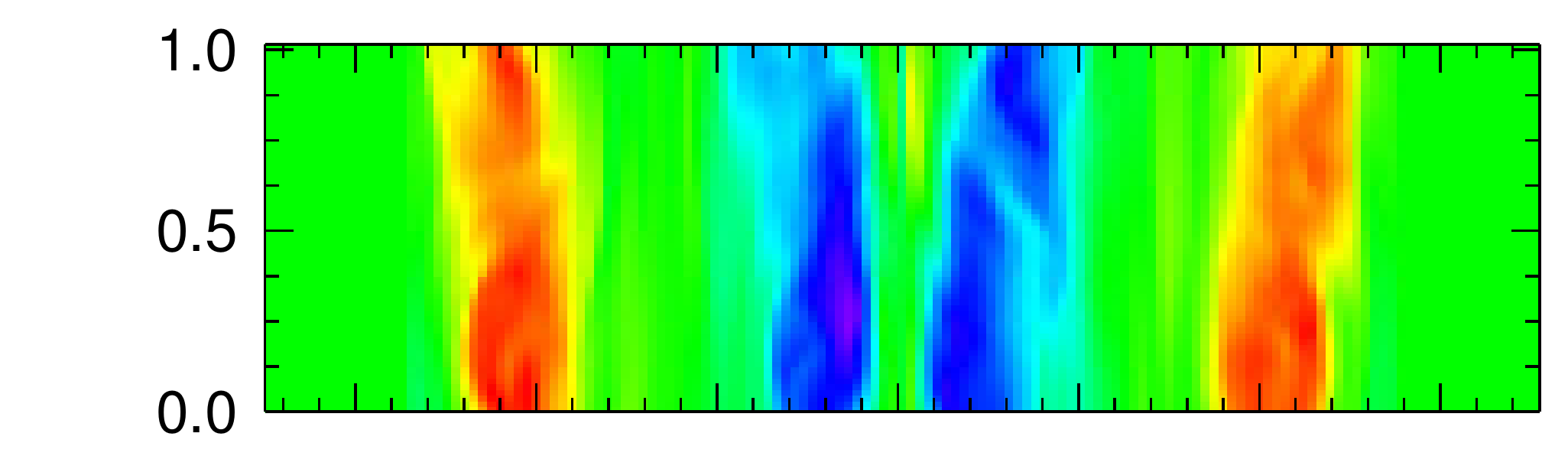}
\centering\includegraphics[scale=0.65, trim=1.0cm 3cm 0.75cm 6cm, clip=true]{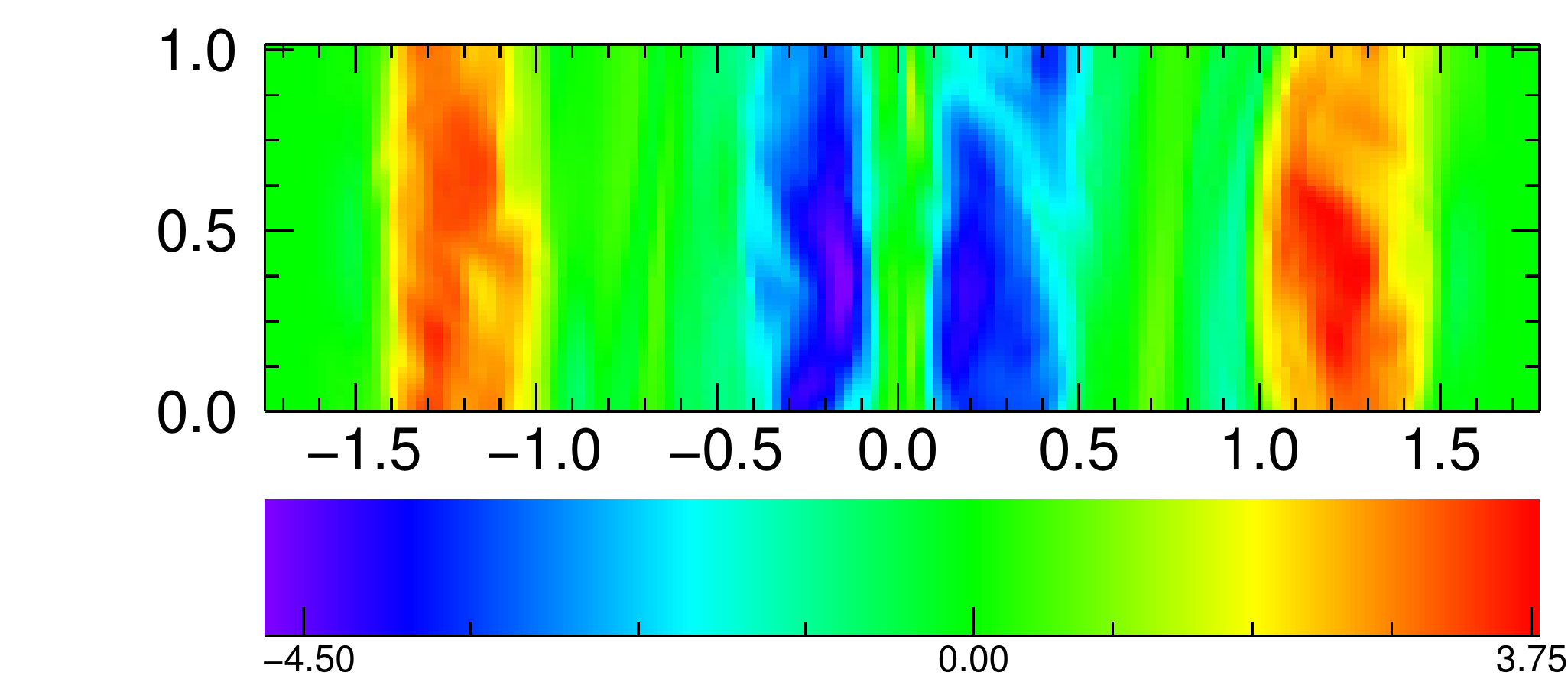}
\caption{Same vertical cuts as in Figure \ref{fig:early} at the end of the simulation. At this late time, the twist flux has condensed to the boundaries of the flux system, with opposite signs at the PIL (outer stripes, red/yellow) and CH (inner stripes, blue/teal).
\label{fig:late}}
\end{figure*}

\begin{figure*}[!p]
\centering\includegraphics[scale=0.55]{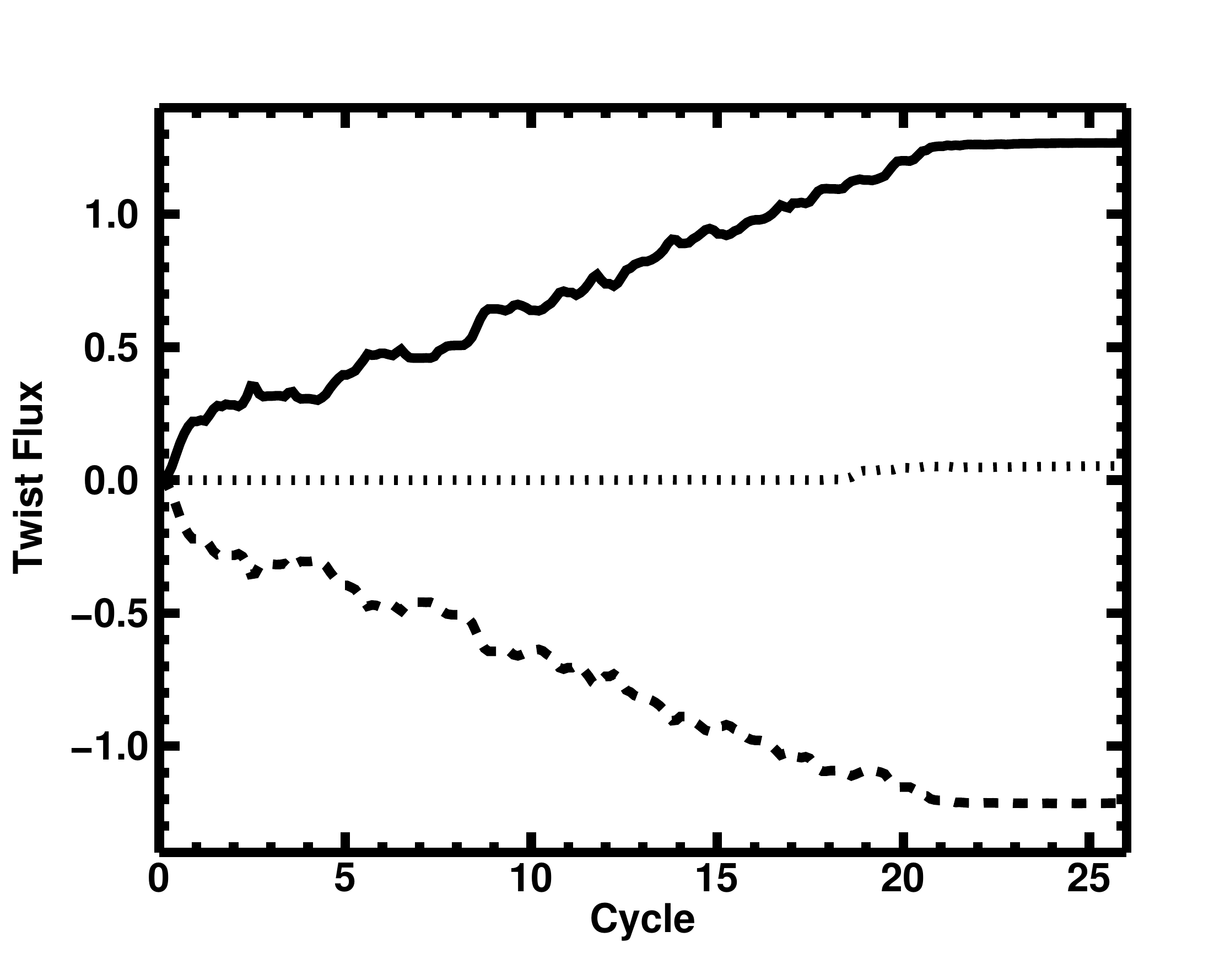}
\caption{Positive and negative twist fluxes, $\Phi_{tw}^+$ and $\Phi_{tw}^-$, from Equation (\ref{twistfluxes}), along the half plane $y=0$, $z\ge0$ ($\phi=0^\circ$). The solid (dashed) curve represents the positive (negative) twist flux, which accumulates at the outer (inner) boundary of the hexagonal flow pattern adjacent to the PIL (CH). The dotted curve is the sum of the two, i.e., the net twist flux through the half plane.  Similar results were obtained on half planes at other angles $\phi$ shown in Figures \ref{fig:early} and \ref{fig:late}.
\label{fig:signedflx}}
\end{figure*}

To a very good approximation, therefore, throughout the simulation the signed twist fluxes obey 
\beg{twistfluxequality}
\begin{split}
\frac{d\Phi_{tw}^-}{dt} &= - \frac{d\Phi_{tw}^+}{dt}, \\
\Phi_{tw}^- &= - \Phi_{tw}^+,
\end{split}
\done
i.e., the fluxes are opposite in sign and equal in magnitude. At early times in Figure \ref{fig:signedflx}, the twist fluxes increase together rapidly; at later times, the oppositely signed twist fields at the interface between the two flux tubes in the $y=0$, $z\ge0$ half plane largely cancel. This cancellation moderates the rise of the accumulated twist fluxes, leaving primarily the residual fluxes that aggregate at the inner and outer boundaries of the hexagonal pattern to contribute to $\Phi_{tw}^-$ and $\Phi_{tw}^+$, respectively. We observed qualitatively similar behaviors of the signed twist fluxes through all the other plane cuts shown in Figures \ref{fig:early} and \ref{fig:late}. Those results will be examined in more detail and analyzed quantitatively in the next subsection. \par

\begin{figure*}[!p]
\centering\includegraphics[scale=0.55]{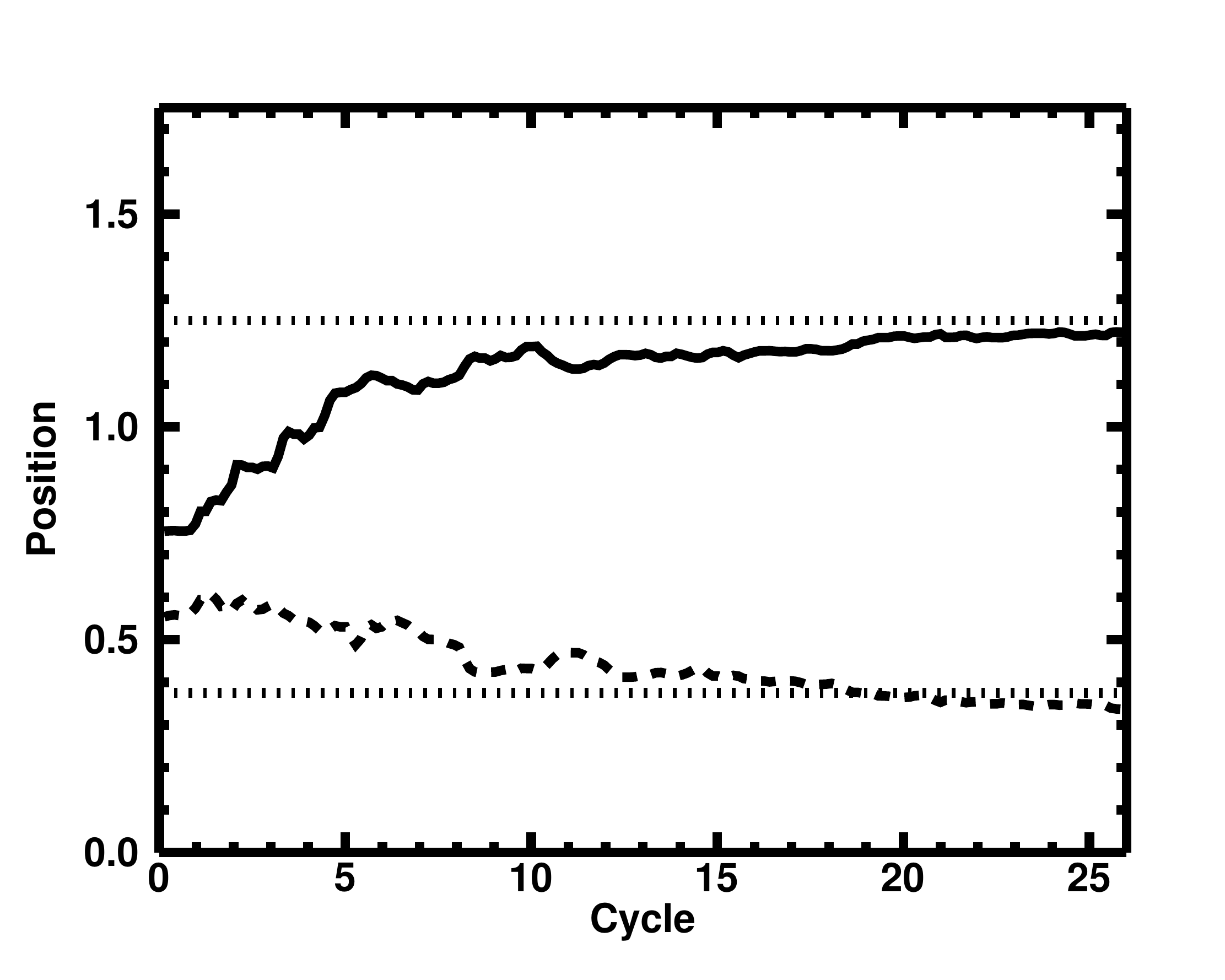}
\caption{Flux-weighted average positions, $\langle s_+ \rangle$ and $\langle s_- \rangle$, from Equation (\ref{weightpos}), of the signed twist fluxes shown in Figure \ref{fig:signedflx}. The solid (dashed) curve represents the position of the positive (negative) twist flux, which accumulates at the outer (inner) boundary of the hexagonal flow pattern adjacent to the PIL (CH). The dotted lines at $s = 3a_0 = 0.375$ and $s = 10a_0 = 1.25$ mark the boundaries of the hexagonal annulus of rotational flows. Similar results were obtained on half planes at other angles $\phi$ shown in Figures \ref{fig:early} and \ref{fig:late}.
\label{fig:weighted}}
\end{figure*}

In Figure \ref{fig:weighted} we plot the flux-weighted positions, $\langle s_+ \rangle$ and $\langle s_- \rangle$, of the signed twist fluxes (\ref{twistfluxes}) in the same half plane $y=0$, $z\ge0$,
\beg{weightpos}
\begin{split}
\langle s_+ \rangle &= \frac{1}{\Phi_{tw}^+} \int_0^{L_x} dx \int_0^{L_z} dz \; {s B_{tw}^+}, \\
\langle s_- \rangle &= \frac{1}{\Phi_{tw}^-} \int_0^{L_x} dx \int_0^{L_z} dz \; {s B_{tw}^-},
\end{split}
\done
where $s = \sqrt{y^2+z^2}$. At early times, the separation of the weighted positions is about $2r_i = 1.6a_0 = 0.20$, very close to the diametric separation of the locations of maximum angular velocity within each rotation cell. The absolute positions agree well with the observed average locations of the thin ribbons of twist flux in the first panel of Figure \ref{fig:early}. As time passes and reconnection enables the inverse cascade of magnetic helicity, on the other hand, the weighted positions of the signed fluxes migrate toward the boundaries of the hexagonal annulus, with the positive (negative) flux approaching the PIL (CH). The inner and outer limits of the hexagonal flow pattern, at radii $3a_0 = 0.375$ and $10a_0 = 1.25$, respectively, are indicated by the dotted lines in Figure \ref{fig:weighted}. Both are close to the calculated flux-weighted positions at late times in the simulation, which converge to approximately 0.30 and 1.20, respectively. Similar behaviors of the weighted positions were observed for the other plane cuts shown in Figures \ref{fig:early} and \ref{fig:late}. \par

\begin{figure*}[!p]
\centering\includegraphics[scale=0.55]{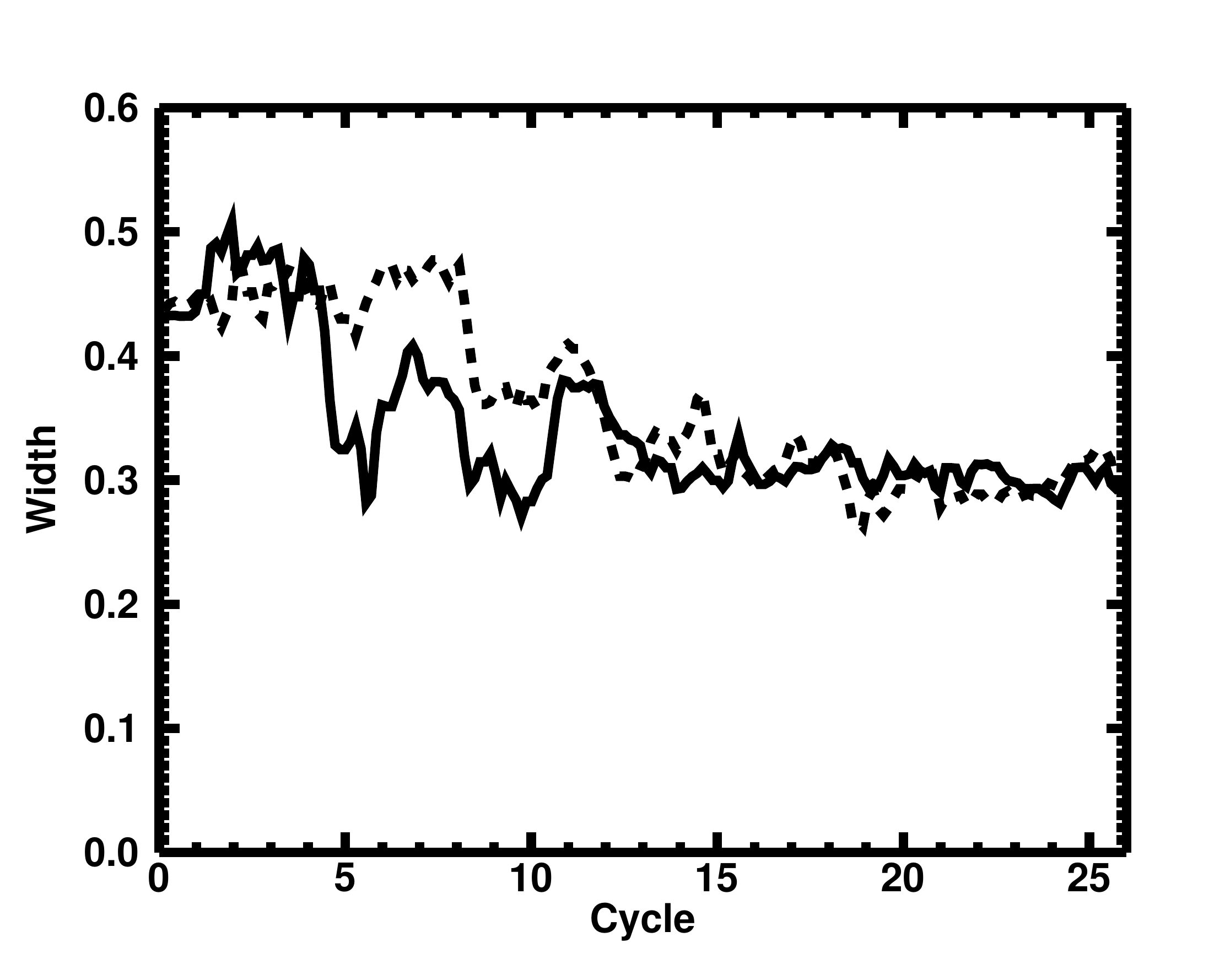}
\caption{Flux-weighted average widths, $\langle w_+ \rangle$ and $\langle w_- \rangle$, from Equation (\ref{weightwidth}), of the signed twist fluxes shown in Figure \ref{fig:signedflx}. The solid (dashed) curve represents the width of the positive (negative) twist flux, which accumulates at the outer (inner) boundary of the hexagonal flow pattern adjacent to the PIL (CH). Similar results were obtained on half planes at other angles $\phi$ shown in Figures \ref{fig:early} and \ref{fig:late}.
\label{fig:widths}}
\end{figure*}

We also calculated the flux-weighted full widths, $\langle w_+ \rangle$ and $\langle w_- \rangle$, of the signed twist fluxes in the half plane, 
\beg{weightwidth}
\begin{split}
\langle w_+ \rangle &= 2 \sqrt{\langle s^2_+ \rangle - \langle s_+ \rangle^2}, \\
\langle w_- \rangle &= 2 \sqrt{\langle s^2_- \rangle - \langle s_- \rangle^2}, \\
\end{split}
\done
where 
\beg{weightpossq}
\begin{split}
\langle s^2_+ \rangle &= \frac{1}{\Phi_{tw}^+} \int_0^{L_x} dx \int_0^{L_z} dz \; {s^2 B_{tw}^+}, \\
\langle s^2_- \rangle &= \frac{1}{\Phi_{tw}^-} \int_0^{L_x} dx \int_0^{L_z} dz \; {s^2 B_{tw}^-}.
\end{split}
\done
The results are shown in Figure \ref{fig:widths}. At early times, the widths reflect the separation between the concentrated bands within the two flux tubes, which is $2\sqrt{3}a_0 = 0.43$ for both the positive and negative fluxes in this $y = $ plane cut.  At late times, the widths measure primarily the radial extent of the condensed bands of twist flux near the PIL and CH, respectively. The late-time widths are essentially equal to each other, and have decreased to about 0.30. Note that, on average, the width decreases steadily with time, but appears to be close to saturation by the end of the 21 rotations. Furthermore, we find a similar behavior for a simulation with only $N = 30$ flux tubes, indicating that the final width of the condensed region is independent of its radial position. Note also that by the end of the simulation, the width approximately equals the diameter of our rotation cells, $2a_0 = 0.25$, as conjectured by \citet{Antiochos13}. We observed similar behaviors of the weighted widths for the other plane cuts shown in Figures \ref{fig:early} and \ref{fig:late}. \par

\subsection{Accumulation Rate}\label{sec:rate}

\begin{figure*}[!p]
\centering\includegraphics[scale=0.55]{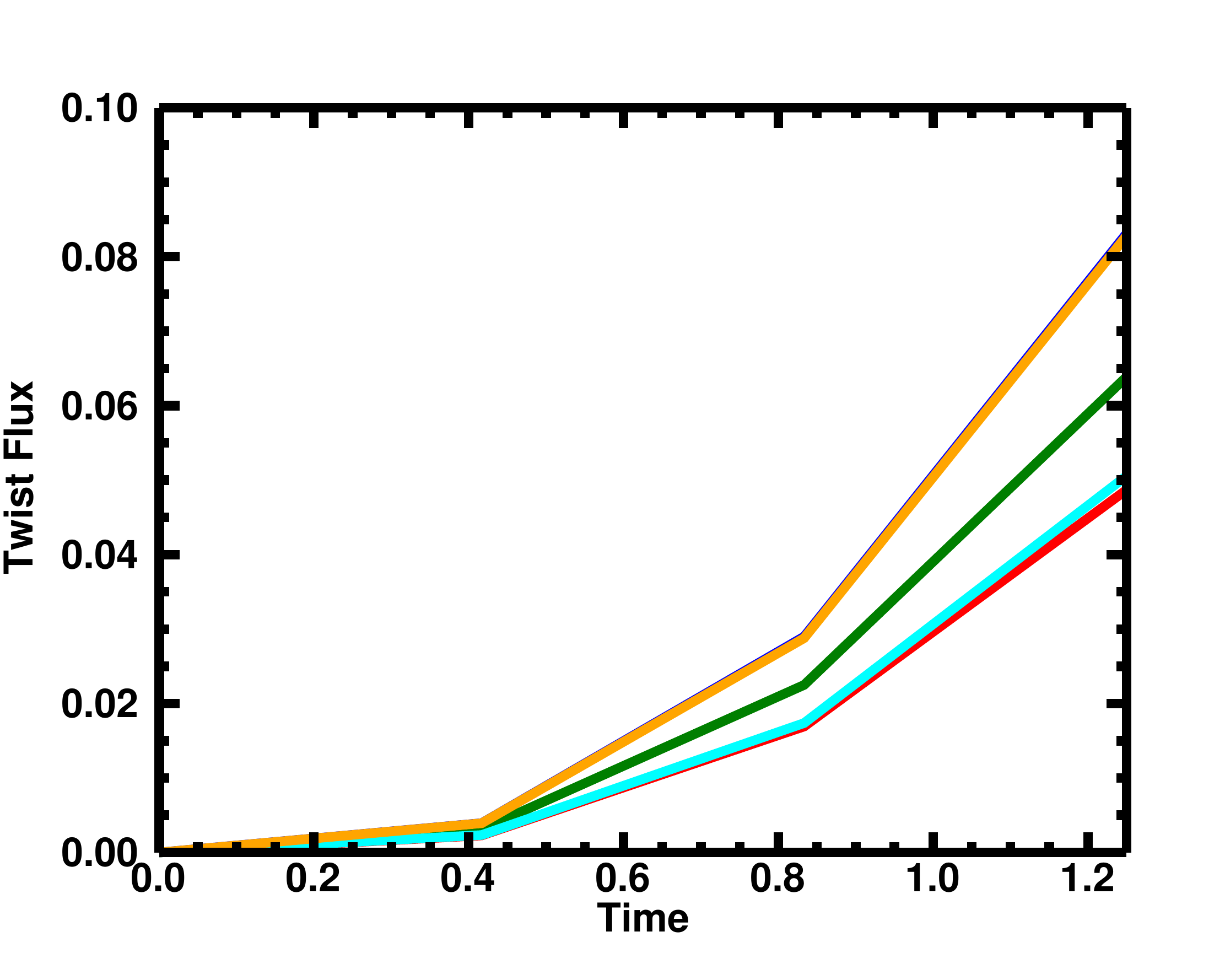}
\caption{Positive twist flux $\Phi_{tw}^+$ during the first third of the first twist cycle through vertical half planes at azimuthal angles $\phi = 0^\circ$ (red), $30^\circ$ (blue), $45^\circ$ (green), $60^\circ$ (cyan), and $90^\circ$ (orange). The individual curves reflect the number of flux tubes cut by the planes early in the simulation.
\label{fig:twistflx1}}
\end{figure*}

In analogy to Equation (\ref{twistfluxes}) for the $\phi=0$ half plane, we calculated the signed twist fluxes versus time for each of the cuts shown in Figures \ref{fig:early} and \ref{fig:late}. Results for the positive twist flux, $\Phi_{tw}^+$, through the half planes early in the first twist cycle are displayed as colored curves in Figure \ref{fig:twistflx1}. The rates of flux increase vary significantly among the different planes, owing to changes in the number of flux tubes that are cut through versus the azimuthal angle $\phi$ (cf.\ Figure \ref{fig:early}). The fluxes along the planes at 30$^\circ$ (blue) and 90$^\circ$ (orange) represent four flux tubes each and rise fastest; the flux along the 0$^\circ$ plane (red) represents two flux tubes and rises slowest, while that along the 60$^\circ$ plane (cyan) is similar; and the flux along the 45$^\circ$ plane (green) rises at an intermediate rate. 
Prior to reconnection setting in, the evolution of the twist flux is perfectly ideal. Inspection of the plane cuts like those in Figure \ref{fig:early} reveals that reconnection of opposite-polarity fluxes from adjacent flux tubes begins to occur at around $t=1.12$ at 30$^\circ$, but does not yet commence at this time at 90$^\circ$. There also is evidence for buckling of some of the flux tubes, leading to partial penetrations of certain planes. Such features can be observed already at time $t=1$ in Figure \ref{fig:early} at 45$^\circ$ and 60$^\circ$. \par

\begin{figure*}[!p]
\centering\includegraphics[scale=0.55]{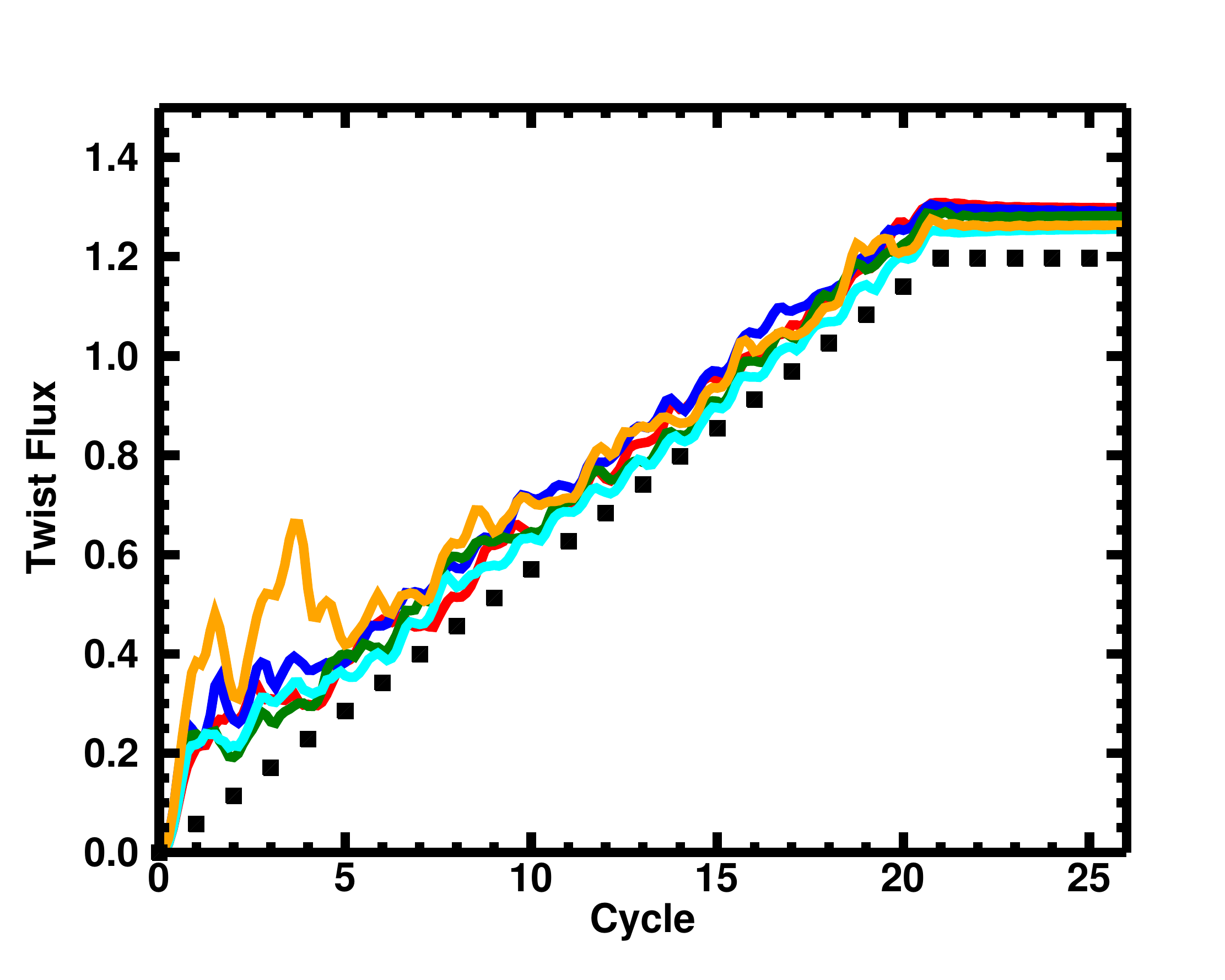}
\caption{Positive twist flux $\Phi_{tw}^+$ versus twist cycle through vertical half planes at azimuthal angles $\phi = 0^\circ$ (red), $30^\circ$ (blue), $45^\circ$ (green), $60^\circ$ (cyan), and $90^\circ$ (orange). Also shown is the accumulated twist flux (filled squares) predicted by Equation (\ref{phitwist}), which is based on the helicity condensation model.
\label{fig:twistflx2}}
\end{figure*}

In contrast to the early evolution of the twist flux shown in Figure \ref{fig:twistflx1}, which depends strongly upon azimuthal angle $\phi$, at late times in the simulation the twist flux is approximately independent of angle. This result, suggested qualitatively by Figure \ref{fig:late}, is confirmed quantitatively in Figure \ref{fig:twistflx2}. Clearly, the flux distribution is becoming more nearly azimuthally symmetric as reconnection transfers the twist fluxes toward the PIL and the CH, where they condense. We showed in \S \ref{sec:helicity} that, if the twist flux is concentrated in discrete bands that enclose a known amount of axial flux, then the helicity injection rate is given by (\ref{dhdttotal}). Here, that relationship takes the explicit form 
\beg{dhdtglobal1}
\begin{split}
\frac{dH}{dt} &= 2 \sum_i \langle \Psi_{ax} \rangle_{r_i} \frac{d\Phi_{tw}(r_i)}{dt} \\
&= 2 \left( \Phi_{CH} \frac{d\Phi_{tw,CH}}{dt} + \Phi_{PIL} \frac{d\Phi_{tw,PIL}}{dt} \right).
\end{split}
\done
Because the twist fluxes are equal in magnitude and opposite in sign, as shown in \S \ref{sec:distribution}, we have 
\beg{dtwdtglobal}
\frac{d\Phi_{tw,CH}}{dt} = - \frac{d\Phi_{tw,PIL}}{dt},
\done
and the preceding relation (\ref{dhdtglobal1}) simplifies to 
\beg{dhdtglobal2}
\frac{dH}{dt} = 2 \left( \Phi_{PIL} - \Phi_{CH} \right) \frac{d\Phi_{tw,PIL}}{dt}.
\done
\par

As described in \S \ref{sec:helicity}, the total helicity injection rate is simply the sum of all of the individual rates within the $N = 84$ flux tubes, so
\beg{HinjTotal}
\frac{dH}{dt} = N {\frac{dH_{f}'}{dt}}.
\done
In addition, the total axial flux in the annular region between the CH and the PIL is approximately the sum of the individual axial fluxes within the $N$ flux tubes that essentially fill the region, so we also have 
\beg{FluxTotal}
\Phi_{PIL} - \Phi_{CH} \approx N \Phi_f,
\done
where $\Phi_f = \pi a_0^2 B_0$ is the total axial magnetic flux per tube. Substituting from the above two expressions into Equation (\ref{dhdtglobal2}), and solving for the twist-flux generation rate, we obtain 
\beg{dphidt}
\frac{d\Phi_{tw,PIL}}{dt} = \frac{1}{2 \Phi_f} \frac{dH_{f}'}{dt}.
\done
Thus, we find that the helicity condensation model predicts that the rate of twist flux accumulation at the boundaries of the flux system equals the effective rate of twist flux generation at the outer edge of an individual flux tube. \par

The helicity injection rate per tube is given by (\ref{dhdtanalytic}), hence, the rate of generation of twist flux at the PIL is 
\beg{dphidt2}
\frac{d\Phi_{tw,PIL}}{dt} \approx \frac{1}{12} \Omega_0 a_0^2 B_0 f(t).
\done
The average twist-flux generation rate and the amount of twist flux generated in one twist cycle are, respectively, 
\beg{avgdphidt}
\begin{split}
\langle \frac{d\Phi_{tw,PIL}}{dt} \rangle &\approx \frac{1}{24} \Omega_0 a_0^2 B_0 \approx 1.7 \times 10^{-2}, \\
\Delta \Phi_{tw,PIL} &\approx \frac{1}{24} \Omega_0 a_0^2 B_0 \tau \approx 5.8 \times 10^{-2}.
\end{split}
\done
These values are somewhat smaller than the realized rate of twist flux generation in the individual flux tubes, given in Equations (\ref{avgdtwdt}) and (\ref{changetw}), by the numerical coefficient 
$15/24=0.625$. This is due to the concentration of the twist flux in the interior of the flux tubes by the assumed angular rotation profile. \par

Now integrating (\ref{dphidt2}) over time, we obtain for the accumulated twist flux
\beg{phitwist}
\Phi_{tw,PIL}(t) = 5.8 \times 10^{-2} \left[t - \frac{\tau}{2\pi} \sin \left(2\pi\frac{t}{\tau}\right) \right].
\done
Therefore, the total twist flux accumulated at the PIL at the end of 21 twist cycles is predicted to be $\Phi_{tw,PIL} = 1.22$. This is in excellent agreement with our numerical results, as shown by the data points (filled squares) from the analytic expression (\ref{phitwist}) included in Figure \ref{fig:twistflx2}. \par

We have further confirmed the prediction of Equation (\ref{dphidt}) that the rate of twist flux accumulation is equivalent to that at the edge of an individual flux tube. A second simulation was performed that included only $N=30$ flux tubes, corresponding to a much smaller polarity region, in which we removed the outermost two layers of rotation cells shown in Figure \ref{fig:initial}. The late-time twist-flux accumulation rate agreed very well with that shown in Figure \ref{fig:twistflx2}. \par

\subsection{Magnetic Shear}\label{sec:shear}

A key indicator of the amounts of free energy and helicity contained in the magnetic field of a filament channel is its shear. Observationally, a standard way to quantify the shear is to measure the angle between a field line of the channel and the PIL. Because our model configuration lacks a true PIL, however, this measurement cannot be made. Instead, we determine the displacement of the footpoints of the magnetic field lines along our equivalent PIL. We adopt this quantity, which is zero for the initial potential field, as our proxy for the magnetic shear. \par

The equation describing the trajectory of a magnetic field line in space is 
\beg{fieldlineeq1}
\frac{d\vecr}{d\ell} = \frac{\vecB}{B},
\done
where $\ell$ denotes the length along the field line. Assuming that the magnetic field is approximately cylindrically symmetric, the field-line equation can be rewritten 
\beg{fieldlineeq2}
\frac{\rho d\phi}{dx} = \frac{B_\phi(\rho)}{B_x(\rho)},
\done
where $\rho$ is the cylindrical radial coordinate. This equation can be integrated immediately to obtain the angular displacement $\Delta \phi$ from one footpoint to the other,
\beg{shearangle1}
\Delta \phi (\rho) = \frac{B_\phi(\rho)}{B_x(\rho)} \frac{L_x}{\rho}.
\done
We evaluate this expression at the center of the band of twist flux at radius $\rho=a$, 
\beg{shearangle2}
\Delta \phi (a) = \frac{B_\phi(a)}{B_x(a)} \frac{L_x}{a},
\done
and recast it into a relationship between the twist and axial fluxes in the band, 
\beg{shearflux1}
\begin{split}
\Phi_{tw} &\approx w L_x B_\phi(a), \\
\Phi_{ax} &\approx 2\pi a w B_x(a),
\end{split}
\done
respectively, where $w$ is the width of the band. The result is 
\beg{shearangle3}
\Delta \phi (a) \approx 2\pi \frac{\Phi_{tw}}{\Phi_{ax}}.
\done
This expression is just the time integral of Equation (\ref{dPhitw2}): the axial flux in the band, $\Phi_{ax}$, is simply the difference between the total enclosed axial fluxes, $\Psi_{ax}$, at the outside and inside of the band, and $\Omega$ is the effective angular rotation rate associated with the angular displacement $\Delta \phi$. \par

\begin{figure*}[!p]
\centering\includegraphics[scale=0.55]{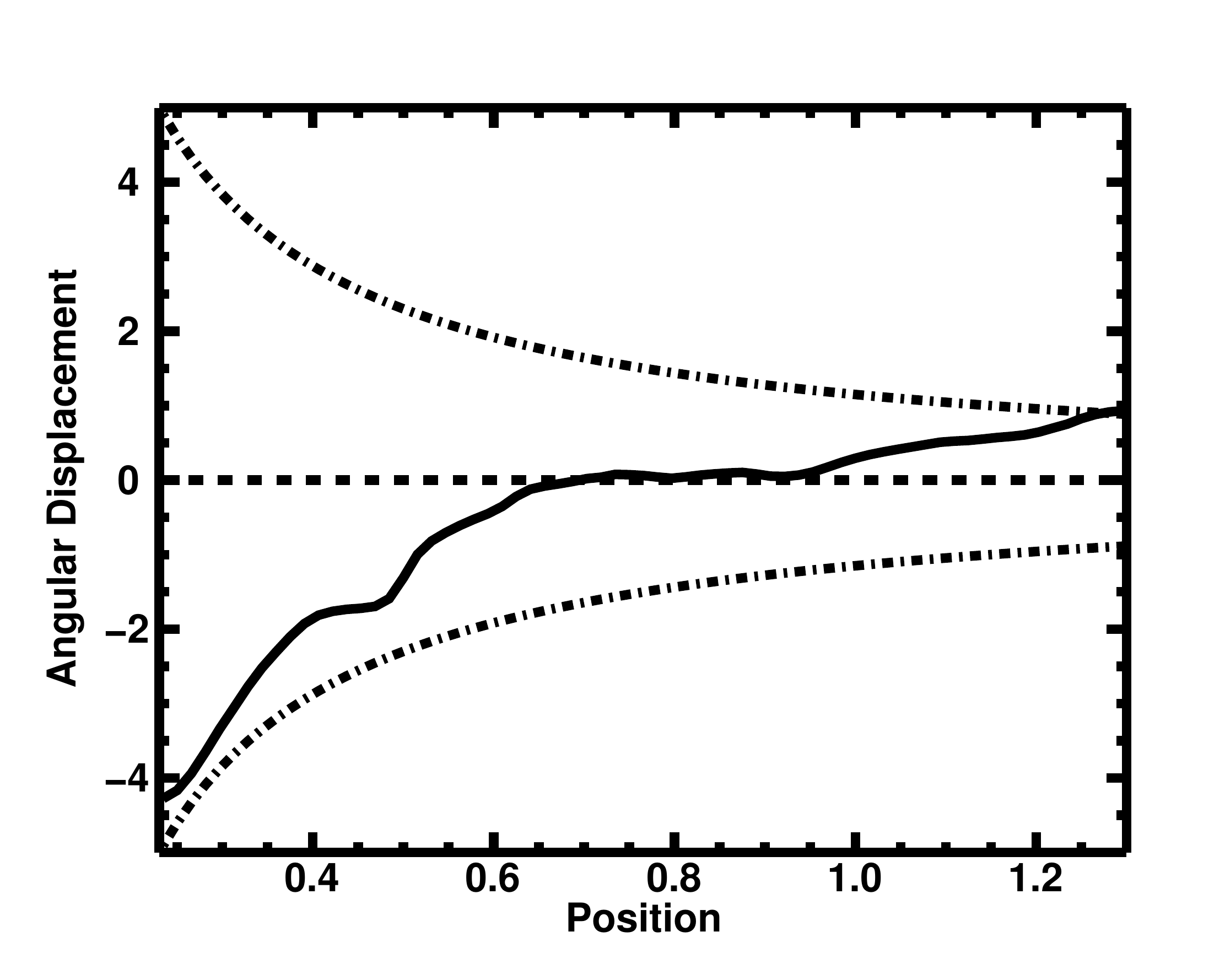}
\caption{The measured shear angular displacement $\Delta \phi$ from Equation (\ref{shearangle1}) in the half plane $y=0$, $z\ge0$ (solid curve), in the region between the two twist bands. Also shown (dot-dash curves) is the predicted displacement from Equation (\ref{shearangle4}), which is based on the helicity condensation model. Untwisted field corresponds to no angular displacement (dashed line).
\label{fig:shearangle}}
\end{figure*}

We can estimate the displacement in (\ref{shearangle3}) by substituting the initial uniform field value, $B_0$, for the axial field strength, $B_x(a)$. After using $B_0=\sqrt{4\pi}$, $\Phi_{tw}=1.22$, 
$w=0.3$, and $L_x=1$ in the above equations, (\ref{shearangle3}) becomes 
\beg{shearangle4}
\Delta \phi (a) \approx \frac{1.15}{a}.
\done
The predicted angular displacements are 0.95 (55$^\circ$) and 3.80 (220$^\circ$) at the PIL and CH ($a$ = 1.2 and 0.3), respectively. Both values agree quite well with the field-line rotations observed in Figure \ref{fig:late}. More generally, we plot as the solid curve in Figure \ref{fig:shearangle} the displacement $\Delta \phi (\rho)$, from Equation (\ref{shearangle1}), in the $y=0, z\ge0$ half plane. For comparison, the prediction from (\ref{shearangle4}) is displayed as the dashed curves for both positive and negative twist fields. The agreement between the predicted and observed shear displacements clearly is very good. \par

\subsection{Application to Filament Channels}\label{sec:channels}


We now use the above results to determine the time scales for filament-channel formation and magnetic-helicity condensation in the solar atmosphere. To do this, we begin by relating our model angular rotation constant $\Omega_0$ to the vorticity $\omega_0$ of supergranulation cells. Equating the vorticity to the average rotation rate of our cells, we find 
\beg{vorticity}
\omega_0 = \frac{2}{a_0^2} \int_0^{a_0} dr \; r \Omega(r) = \frac{2}{15} \Omega_0,
\done
after integrating Equation (\ref{omega}) for the rotation profile. Substitution into (\ref{dtwdtanalytic}) gives for the peak rate of twist-flux generation 
\beg{dtwdtchannel1}
\frac{d\Phi_{tw}}{dt} = \omega_0 a_0^2 B_0.
\done
Taking the scale to be that of a typical supergranule radius, $a_0 = 1.4 \times 10^9$ cm, and the angular velocity to be set by typical supergranular velocities, $\omega_0 a_0  = 5 \times 10^{4}$ cm s$^{-1}$, we find that 
\beg{dtwdtchannel2}
\frac{d\Phi_{tw}}{dt} = 7 \times 10^{13} B_0 \; {\rm Mx\; s}^{-1}.
\done
If we assume that the twist flux is generated in a semicircular volume of width $w = 2a_0$, then the rate of accumulation of twist field is 
\beg{dbdtchannel}
\frac{dB_{tw}}{dt} = \frac{2}{\pi a_0^2} \frac{d\Phi_{tw}}{dt} = \frac{2}{\pi} \omega_0 B_0.
\done
Defining the channel formation time, $\tau_{fc}$, to be the time required for the twist field $B_{tw}$ to acquire the same strength as the vertical field $B_0$, we obtain 
\beg{tchannel}
\tau_{fc} = \frac{\pi}{2} \omega_0^{-1} = 4.5 \times 10^4 \; {\rm s},
\done
or approximately half a day. This is more than fast enough to explain the formation and maintenance of filament channels \citep[e.g.,][]{Anderson05}. It should be emphasized, however, that this estimate is most likely a lower limit on the formation time, because we assumed above that all supergranules are injecting the same sense of twist into the corona. In reality, there is only a hemispheric preference, so some non-negligible fraction is injecting the opposite sense of twist, which would slow down the rate of shear buildup. \par

The peak rate of magnetic helicity injection per supergranule, given by one-half of Equation (\ref{dhdtanalytic}), becomes 
\beg{dhdtsg1}
\frac{dH_{sg}}{dt} = \frac{5\pi}{8} \omega_0 a_0^4 B_0^2
\done
after using (\ref{vorticity}) for the vorticity. Substituting for the fixed parameters, we obtain the numerical value 
\beg{dhdtsg2}
\frac{dH_{sg}}{dt} = 2.7 \times 10^{32} B_0^2 \; {\rm Mx}^2 \; {\rm s}^{-1}.
\done
Assuming that 50$\%$ of the solar surface is covered by closed magnetic flux where the injected helicity can be stored, we obtain for the full-Sun rate of helicity injection 
\beg{dhdtsol}
\frac{dH_{\odot}}{dt} = 2 \frac{R_{\odot}^2}{a_0^2} \frac{dH_{sg}}{dt} = 1.4 \times 10^{36} B_0^2 \; {\rm Mx}^2 \; {\rm s}^{-1}.
\done
Over the duration $3.5 \times 10^8$ s of the sunspot cycle, the total injected helicity is 
\beg{dhsol1}
\Delta H_{\odot} \approx 3.5 \times 10^{44} B_0^2 \; {\rm Mx}^2.
\done
Assuming an average flux density $B_0 = 10$ G \citep{1993PhDT.......225H}, we find for the total helicity injected over the cycle 
\beg{dhsol2}
\Delta H_{\odot} \approx 3.5 \times 10^{46} \; {\rm Mx}^2,
\done
which agrees well with the estimate by \citet{Antiochos13}. This value is similar to estimates of the total magnetic helicity expelled by the corona in the solar wind and as coronal mass ejections, and stored in the corona by the Sun's differential rotation \citep{DeVore00b}. \par

\section{Discussion}\label{sec:discussion}

In this paper, we modeled the injection and transport of magnetic helicity in a plane-parallel model of the solar corona \citep{Parker72} using helicity-conserving numerical simulations. Helicity was injected into the corona at the photospheric boundary by numerous rotation cells, emulating the solar supergranulation, and then transported by magnetic reconnection throughout the coronal volume where the magnetic field was twisted by the imposed rotational motions. We found that the reconnection cancelled the opposite-polarity twist fluxes from adjoining flux tubes twisted at the scale of the individual rotation cells, leaving a large-scale residual twist flux that condensed at the inner and outer boundaries of the region of imposed photospheric flows. These boundaries constituted the coronal hole (CH) and polarity inversion line (PIL) of our model system, and the twist fluxes condensing at the two locations were opposite in sign. All of these qualitative features of the simulation results agree well with the helicity condensation model developed by \citet{Antiochos13}. \par

The simple initial and boundary conditions imposed in our simulations also enabled us to quantify certain aspects of our results. Owing to the uniformity of the magnetic field and the rotation speeds in the model, the twist fluxes condensing at the CH and PIL were equal in magnitude, as well as opposite in sign (\S \ref{sec:distribution}). This leads to the prediction that the rate of accumulation of twist flux, into two bands localized at the CH and PIL boundaries, is equivalent to the rate of accumulation at the outer edge of a single rotation cell (\S \ref{sec:rate}). We calculated the latter exactly and used it to verify the simulation data. The resulting expression for the twist flux accumulation rate was converted into a prediction for the shear angular displacement of field lines rooted near the CH and PIL boundaries (\S \ref{sec:shear}), which we confirmed by comparing it with measured values from the simulation. We emphasize that all of these demonstrations rely heavily upon the conservation of magnetic helicity by our high-fidelity numerical-simulation model, ARMS \citep[e.g.,][]{DeVore08}. \par

Our results have important implications for filament-channel formation on the Sun. The simulations confirm that the inverse cascade of magnetic helicity from small to large scales indeed occurs, as has been found in other magnetohydrodynamic simulations generally \citep{1993noma.book.....B} and postulated for the helicity condensation model specifically \citep{Antiochos13}. This cascade culminates in the twist flux accumulating at the PIL, yielding a global magnetic shear concentrated at the location where filament channels are known to form \citep{Martin98,Gaizauskas00}. The quantitative application of our analytical and numerical results to filament channel formation (\S \ref{sec:channels}) yields numbers for the accumulation of twist flux, helicity, and magnetic shear that are generally in accord with solar observations. \par

The simplified geometry of our system also caused twist flux to condense at the CH boundary, where it accumulated much like the twist flux at our PIL. Unlike coronal holes on the Sun, the field lines in our model CH are line-tied at both ends -- therefore, they are effectively closed, rather than open -- and are not free to expand outward but compress together at the center of our domain. These artificial constraints imposed by our simulation geometry can be alleviated by adopting a truly bipolar magnetic field above a single planar photosphere with an open top boundary. This extension of our current modeling is underway. The results presented here nevertheless demonstrate conclusively that, as predicted by \citet{Antiochos13}, the twist flux condensing at the CH boundary is opposite in sign to that at the PIL. Furthermore, it is opposite in sign to that generated by the rotational motions that are present in the center of solar coronal holes (but are omitted from the CH in our current study). Our follow-up simulations, therefore, will test rigorously the prediction of the helicity condensation model that the flux of magnetic helicity at the boundary of a coronal hole, carried away by the slow solar wind, should be of opposite sign to the helicity flux from the interior of the hole, carried away by the fast solar wind \citep{Antiochos13}. \par

Adopting a realistic magnetic-field geometry with a true PIL also will introduce photospheric gradients of the vertical magnetic field into our system. The uniformity of the vertical field and of the rotation speeds in our current study implies that all of the twisted flux tubes are formed with equal twist fluxes. In a nonuniform vertical field, or in a region where the rotation speeds are nonuniform, this equality of the twist fluxes on adjacent tubes will break down. The resulting imperfect cancellation of adjacent twist fluxes is anticipated to lead to a broader distribution of the twist flux condensing at the PIL, compared to that formed in the simulations of this paper. Thus, an investigation of the effect of gradients in the background vertical magnetic field, as well as other factors influencing the distribution of the filament channel flux -- such as the size of the rotation cells at the photosphere -- will be a part of our future studies. \par

The primary conclusion to be drawn from this paper is that the helicity condensation model developed by \citet{Antiochos13} is in excellent agreement with the results of our numerical experiments, both qualitatively and quantitatively. This is a very encouraging, albeit far from final, step forward toward demonstrating that the helicity condensation model explains the heretofore mysterious origin of solar filament channels. \par



\acknowledgements

\begin{figure*}[!p]
\includegraphics[scale=0.55]{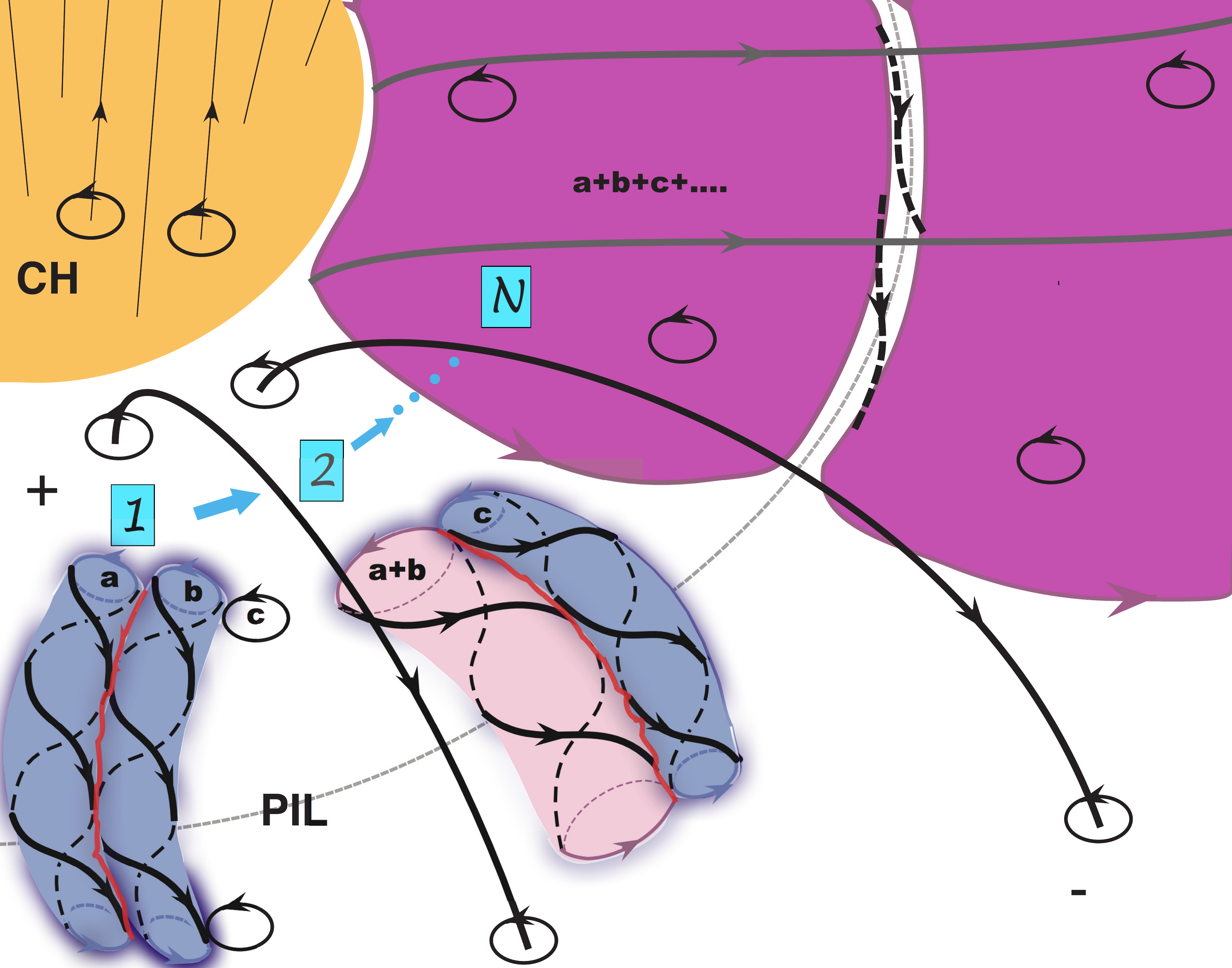}
\caption{Helicity condensation model, after \citet{Antiochos13}. The yellow region represents the photosphere within a coronal hole (`CH'). The thin gray dashed line is a polarity inversion line (`PIL'). The counter-clockwise circular arrows represent a few individual rotation cells among the whole, densely distributed population of such cells on the surface of the Sun. Thick black curves represent magnetic field lines, drawn solid on the top side and dashed on the underside of the magnetic flux tubes to which they belong. Yellow curves indicate contact points where the azimuthal fields of adjacent tubes are antiparallel and so can reconnect. Individual flux tubes are colored light blue (`a', `b', and `c') or light purple (`a+b'). Step numbers (`1', `2', `$N$') indicate the progressive transport of azimuthal magnetic flux from the injection scale of the rotation cells to the global scales of the CH and PIL. Ultimately, the azimuthal flux either condenses at the PIL to form a filament channel (thick black dashed lines linking the dark purple regions) or is released into the solar wind after propagating into the open field at the CH boundary.
\label{fig:condensation}}
\end{figure*}

\begin{figure*}[!p]
\centering\includegraphics[scale=0.75,trim=0cm 8.0cm 0cm 8.0cm, clip=true]{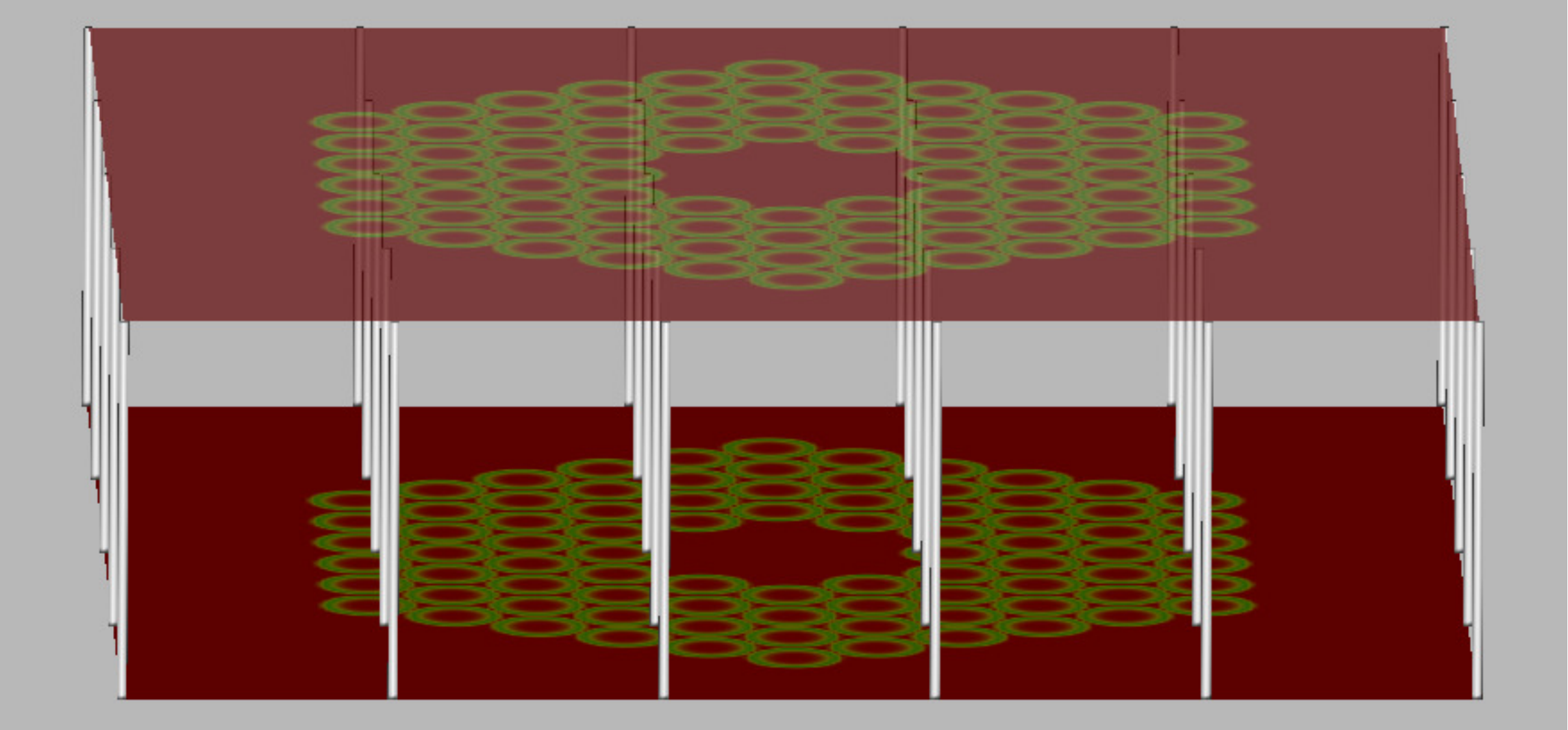}
\centering\includegraphics[scale=0.67]{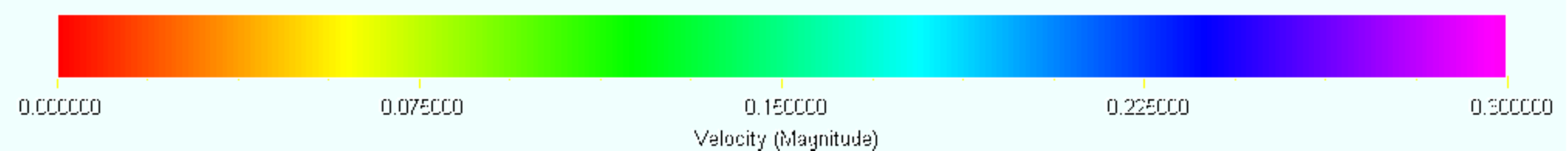}
\caption{Initial simulation setup having a uniform magnetic field between two plates and 84 rotation cells -- locations where the field is twisted by line-tied flows -- in a hexagonal array on each plate. White lines are magnetic field lines. Color shading on the top and bottom planes indicates the magnitude of the flow velocity: red shading indicates locations where the flow speed is zero, green where it is nonzero. The outer boundary of the flow region represents a PIL; the inner flow-free region represents a CH (compare with Figure \ref{fig:condensation}).
\label{fig:initial}}
\end{figure*}

We thank K.\ Dalmasse, S.\ Guidoni, A.\ Higginson, S.\ Masson, and K.\ Muglach for illuminating discussions and helpful suggestions, and we greatly appreciate J.\ Karpen's contribution of the artwork shown in Figure \ref{fig:condensation}. KJK acknowledges support for this work by a cooperative agreement between the Catholic University of America and NASA Goddard Space Flight Center, as well as additional funding received through the NASA Earth and Space Science Fellowship program. The participation of SKA and CRD was sponsored by NASA's Heliophysics Theory and Supporting Research programs. The numerical simulations were supported by a grant of NASA High End Computing resources to CRD.

\bibliography{helicity}

\newpage

\end{document}